\documentclass[twocolumn,floatfix]{revtex4-2}%
\usepackage[dvipdfmx]{graphicx}%
\usepackage{amsmath}%
\usepackage{amsfonts}%
\usepackage{amssymb}
\usepackage{bm}
\usepackage[dvipdfmx]{color}

\def\d{{\partial}}
\def\s{{\sigma}}
\def\e{{\epsilon}}
\def\k{{ {\bm k} }}

\def\q{{ {\bm q} }}

\def\0{{ {\bm 0} }}
\def\w{{\omega}}
\def\a{{\alpha}}
\def\b{{\beta}}

\allowdisplaybreaks[4]

\begin{document}
\title{\textcolor{black}{
Impact of multiband effects on non-Fermi-liquid transport phenomena in bilayer nickelates}
}
\author{
Seiichiro Onari$^{1}$, Daisuke Inoue$^{1}$, Rina Tazai$^{2}$, Youichi
Yamakawa$^{1}$, and Hiroshi Kontani$^1$
}
\date{\today }

 \begin{abstract}
Recently discovered high-$T_c$ superconductivity in thin-film bilayer nickelates La$_3$Ni$_2$O$_7$ under ambient pressure has attracted great interest. 
Non-Fermi-liquid transport behaviors, such as $T$-linear resistivity and a positive Hall coefficient that increases at low temperatures, have been reported in this system.
In this study, we analyze the non-Fermi-liquid transport phenomena in the thin-film bilayer nickelate La$_3$Ni$_2$O$_7$ using a multiorbital tight-binding model.
  In La$_3$Ni$_2$O$_7$, the cold spots composed of Ni $d_{x^2-y^2}$ orbital emerge, since the spin fluctuations cause stronger quasiparticle damping $\gamma$ in the Ni $d_{z^2}$ orbital.
 Notably, in the present study, we derive a rigorous formula for the Hall coefficient $R_H$ incorporating the $\gamma$ in the quasi-quantum metric (qQM) term. We find that the $T$ dependence of $\gamma$ in the qQM term is important in determining $R_H$. In La$_3$Ni$_2$O$_7$, the $T$ dependence of $R_H$ becomes pronounced due to the competition between the positive contribution from the hole band and the negative contribution from the electron band. Moreover, the qQM term plays an important role in describing the Nernst coefficient and other transport phenomena involving the second derivative velocity $v^{\mu\nu}$.
 \end{abstract}
 
\address{
$^1$Department of Physics, Nagoya University,
Nagoya 464-8602, Japan \\
$^2$RIKEN Center for Emergent Matter Science, Wako, Saitama 351-0198, Japan
} 
\sloppy

\maketitle

%%%%%%%%%%%%%%%%%%
%{\it Introduction:} \ 
%%%%%%%%%%%%%%%%%%
\section{INTRODUCTION}

The recent discovery of high-$T_{\rm c}$ superconductivity in thin-film
La$_3$Ni$_2$O$_7$ under ambient pressure has generated significant interest. 
The realized superconducting (SC) transition temperature ($T_{\rm c}$)
is about 40K \cite{Ko-Ni-SC,
Zhou-Ni-SC,Osada-Ni-SC,Ni-thin-trans}. The phase diagram of this system has not been clear
yet. ARPES measurements propose $\sim 20$\% hole-doping for each Ni
atom, and the epitaxial strain is introduced in the thin-film
system \cite{Ni-thin-ARPES,Ni-thin-ARPES2}.
On the other hand, in bulk samples, superconductivity with $T_c \sim 80$K
under $P\sim10$GPa has been reported \cite{Sun-Ni-SC,
Wang-Ni-SC, Zhang-Ni-SC}.
At ambient pressure of the bulk sample, double-stripe spin density wave (SDW) with ${\bm
q}_{\mathrm{s}}\approx(\pi/2,\pm\pi/2)$ appears at
$T_{\mathrm{sdw}}\approx150$K \cite{Dan-Ni-NMR,Mukuda-Ni-NMR,Chen-Ni-mu+SR,Khasanov-Ni-muSR,Chen-Ni-RIXS,Gupta-Ni-RSXS,Ren-Ni-X-ray,NQR-Okayama}.
Furthermore, the charge density wave (CDW) orders have been observed at
the CDW transition temperature $T_{\rm cdw}$ comparable to $T_{\rm sdw}$
\cite{Mukuda-Ni-NMR,Chen-Ni-mu+SR,Khasanov-Ni-muSR,Chen-Ni-RIXS,Gupta-Ni-RSXS,Ren-Ni-X-ray,NQR-Okayama,Xie-Ni-neutron,Liu-Ni-resistivity,Wu-Ni-DW}.

To understand the mechanism of the SC state, it is important to elucidate the parent normal states. 
The non-Fermi-liquid transport phenomena are evidence of strongly correlated electron systems.
In proximity to the quantum critical point of the SDW or CDW, non-Fermi-liquid transport phenomena have been observed in infinite-layer nickelates, Fe-based, and cuprate superconductors
 \cite{Ni-Non-Fermi1,Ni-Non-Fermi2,Ni-Non-Fermi3,Non-Fermi-Cu,Cu-trans1,Cu-trans2,Non-Fermi-Cu-RH,Non-Fermi-Cu-RH2,Non-Fermi,Non-Fermi2,Non-Fermi-FeSeS,Non-Fermi-Fe}.
Very recently, in thin-film bilayer nickelate, the non-Fermi-liquid behaviors of the resistivity $\rho\propto T$ and the positive Hall coefficient $R_H$ that increases at low temperatures have been reported \cite{Ko-Ni-SC,Ni-thin-trans}.

The relaxation-time approximation (RTA) is useful to discuss the quasiparticle transport phenomena in metals \cite{Ziman}.
In the RTA, the relaxation time is given by the inverse of the quasiparticle damping $\gamma_{b,\k}$,  which is proportional to the imaginary part of the self-energy.
Here, $b$ is the Fermi surface, and $\k$ is the momentum index.
The RTA is applicable to various moderately correlated metals, except for systems close to the quantum critical points, like cuprate and Fe-based superconductors \cite{Non-Fermi-Cu,Cu-trans1,Cu-trans2,Non-Fermi-Cu-RH,Non-Fermi-Cu-RH2,Non-Fermi,Non-Fermi2,Non-Fermi-FeSeS,Non-Fermi-Fe}.

In multiband systems, however, the quantum metric (QM) and quasi-QM (qQM) can be significant for various quantum transport phenomena \cite{QM1,QM2,QM3,QM4,QM5,QM6,Tazai-EMCHA}.
 The qQM and QM are the paramount concepts in condensed matter physics. However, as far as we know, the significant effects of the qQM on quasiparticle transport, such as the ordinary Hall effect, have not yet been studied in strongly correlated multiband systems like La$_3$N$_2$O$_7$.

Previous theoretical studies on bulk and thin-film La$_3$Ni$_2$O$_7$ have proposed spin-fluctuation-mediated SC states with sign-reversing gaps, such as $d$-wave and $s{\pm}$-wave states \cite{Lechermann-Ni-SC,Liu-Ni-SC2,Heier-Ni-SC,Liu-Ni-SC,Dagotto-Ni-SC,Yang-Ni-SC,Jiang-Ni-SC,Gu-Ni-SC,Dagotto-Ni-SC2,Sakakibara-Ni-SC,Zhan-Ni-SC,Xue-Ni-SC,Tian-Ni-SC,Lu-Ni-SC,film-Shao-RPA,film-CDMFT} have been proposed.
The normal and SC states have also been studied by employing strong coupling approaches \cite{Lechermann-Ni-SC,Tian-Ni-SC,Christiansson-Ni-GW_EDMFT,Ouyang-Ni-DMFT,Ryee-Ni-SC,Duan-Ni-SC,Qiu-Ni-SC}.
Importantly, we have recently revealed that both inter-layer and intra-layer CDW instabilities emerge concurrently with the SDW instability.
This CDW can be explained by the paramagnon-interference (PMI) mechanism \cite{bilayer-Inoue}.
However, the non-Fermi-liquid transport phenomena remain unsolved. A novel theory is required to address this issue.

In this study, we study the origin of non-Fermi-liquid transport phenomena in
thin-film bilayer nickelate La$_3$Ni$_2$O$_7$ by focusing on many-body
  effects on the QM within a 2D multiorbital
  tight-binding model.
  In La$_3$Ni$_2$O$_7$, the $d_{x^2-y^2}$-orbital cold spots
  emerge due to the orbital dependence of $\gamma$, which is
  driven by the spin fluctuations. For simplicity, we ignore
  the CDW fluctuations.
We derive a rigorous formula for the $R_H$ that incorporates the many-body effects on the qQM, which is a multi-band effect.
We find that the $T$ dependence of qQM term becomes significant in strongly correlated multiband systems with strong $T$-dependent $\gamma_b\gg T$. The $T$ dependence of qQM is not captured within the RTA. We also find that the $T$ dependence of $R_H$ in La$_3$Ni$_2$O$_7$ is pronounced due to the competition between the positive contribution from the hole band and the negative contribution from the electron band.
The obtained $\rho\propto T$ and the enhancement of positive $R_H$ at low
$T$ are consistent with experimental observations
\cite{Ko-Ni-SC,Ni-thin-trans}. We also obtain the negative Nernst and
Seebeck coefficients, which have not yet been experimentally observed in thin-film La$_3$Ni$_2$O$_7$.

\textcolor{black}{
\section{Rigorous formula for the normal Hall coefficient with qQM using the green-function method}}
The aim of this paper is to analyze the quasiparticle transport phenomena in strongly correlated multiband La$_3$Ni$_2$O$_7$.
\textcolor{black}{ In Ref. \cite{Fukuyama}, the $\tau^2$ term of the Hall conductivity $\sigma_{xy}$ was derived within the framework of the Green-function method for single-band models. Subsequently, Refs. \cite{Ogata1,Ogata2} derived the $\tau^2$ term of $\sigma_{xy}$ for multiband models satisfying $\partial_x \partial_y \hat{h}_{\k} = 0$, where $\d^\nu\equiv\frac{\d}{\d k_\nu}$ and $\hat{h}_{\k}$ denotes the Hamiltonian matrix. In the present study, we derive a rigorous formula for the $\tau^2$ term of $\sigma_{xy}$ applicable to general multiband models with $\partial_x \partial_y \hat{h}_{\k} \ne 0$, and apply it to nickelates.}
Based on the formula, we can accurately evaluate the ordinary Hall effect, including the contribution from the qQM, which represents an interband effect.
Importantly, the present formula allows us to evaluate the self-energy effects on the qQM, which give rise to the pronounced temperature dependence of the Hall coefficient in strongly correlated La$_3$Ni$_2$O$_7$, as we will demonstrate in Sec. V.
The qQM $g_a^{\mu\nu}(\e)$ is defined as
\begin{eqnarray}
g_a^{\mu\nu}(\e)&=&
\frac12 \sum_{b\ne
a}[v^{\mu}_{ab}v^{\nu}_{ba}+v^{\nu}_{ab}v^{\mu}_{ba}]{\rm Re}G^{\rm
R}_{b}(\e),\label{eqn:qqm} \\ 
{\rm Re}G^{\rm R}_{b}(\e)&=&\frac{\e-\e_b}{(\e-\e_b)^2+\gamma_b^2(\e)}.
\label{eqn:ReG}
\end{eqnarray}
Here,  ${\hat v}_\k^\nu\equiv \d^\nu {\hat h}_\k$, where $\hat{h}_{\k}$ is the Hamiltonian matrix in the orbital representation. ${v}_{ab,\k}^{\mu} \equiv \langle u_{a,\k}|{\hat {v}}_\k^{\mu}|u_{b,\k}\rangle$ is represented in band basis    ($a,b$).
Importantly, $g_a^{\mu\nu}(\e)$ depends on the temperature through $\gamma_{b,\k}$, which plays a crucial role in strongly correlated electron systems. 
This qQM is similar to the QM given by
\begin{equation}
G_a^{\mu\nu}(\e)=
\sum_{b\ne
a}[v^{\mu}_{ab}v^{\nu}_{ba}+v^{\nu}_{ab}v^{\mu}_{ba}]\frac{1}{(\e-\e_b)^2}.
\end{equation} 

\textcolor{black}{In previous studies on weakly correlated multiband systems \cite{Ogata1,Ogata2}, the effects of the qQM term have been taken into account.} \textcolor{black}{However, in these studies $\gamma_{b,\k}$ was assumed to be a constant $\gamma$, representing disorder effects. Therefore, the $T$ dependence of $R_H$ arising from the strong $T$-dependent qQM term in strongly correlated electron systems, which is central to the present study, has not been considered.} \textcolor{black}{We note that a formula for $\sigma_{xy}$ applicable to multiband systems with $\partial_x\partial_y \hat{h}_{\k}\ne0$ was independently derived in recent study \cite{Ogata-Tsuji}.}

In Ref. \cite{Tazai-EMCHA}, three of the present authors (R.T., Y.Y., and H.K.) have revealed a significant role of the QM in the electronic magnetochiral anisotropy (eMChA), a nonreciprocal transport phenomenon of order $E^2B$ in the electric and magnetic fields.
In the loop-current phase in kagome metals, the derived eMChA conductivity takes a huge value proportional to $\tau^3$, where $\tau$ is the lifetime of electrons.
Significantly, the eMChA is strongly magnified ($\sim 100$ times) by the QM term in multiorbital kagome metals, giving rise to the giant eMChA discovered in kagome metals \cite{Moll}.
This study demonstrates the significance of the QM term in multiorbital systems.

In the following, we study the multiorbital tight-binding model:

\begin{eqnarray}
H=\sum_{i\a,j\b,\s} t_{i\a,j\b}c_{i\a,\s}^\dagger c_{j\b,\s},
\label{eqn:Ham1}
\end{eqnarray}
where $i,j$ represent the indices of the unit-cell ($i=1\sim N$), 
and $\a,\b$ represent the site and/or orbital in each unit-cell ($\a=1\sim M$).
$\s$ is the spin index. In the present La$_3$Ni$_2$O$_7$ model, $M=4$.
In the momentum space, Eq. (\ref{eqn:Ham1}) is rewritten as
\begin{eqnarray}
H=\sum_{\k,\a,\b,\s} h_{\k}^{\a,\b}c_{\k\a,\s}^\dagger c_{\k\b,\s}.
\label{eqn:Ham2}
\end{eqnarray}
Here, $\hat{h}_{\k}$ is the Hamiltonian matrix in the orbital representation.
The eigenvalue equation of $\hat{h}_{\k}$ is
\begin{eqnarray}
{\hat h}_{\k}|u_{a,\k}\rangle = \e_{a,\k}|u_{a,\k}\rangle,
\label{eqn:Bloch}
\end{eqnarray}
where $\e_{a,\k}$ is the $a$-th band-energy ($a=1\sim M$), and $|u_{a,\k}\rangle$ represents the Bloch function.
Hereafter, we will refer to the conduction band as $a$ and the valence band as $b$.

%%%%%%%%%%%%%%%%%%%%%%%%%%%%%%%%%%%%%%%
\begin{figure}[htb]
\includegraphics[width=.99\linewidth]{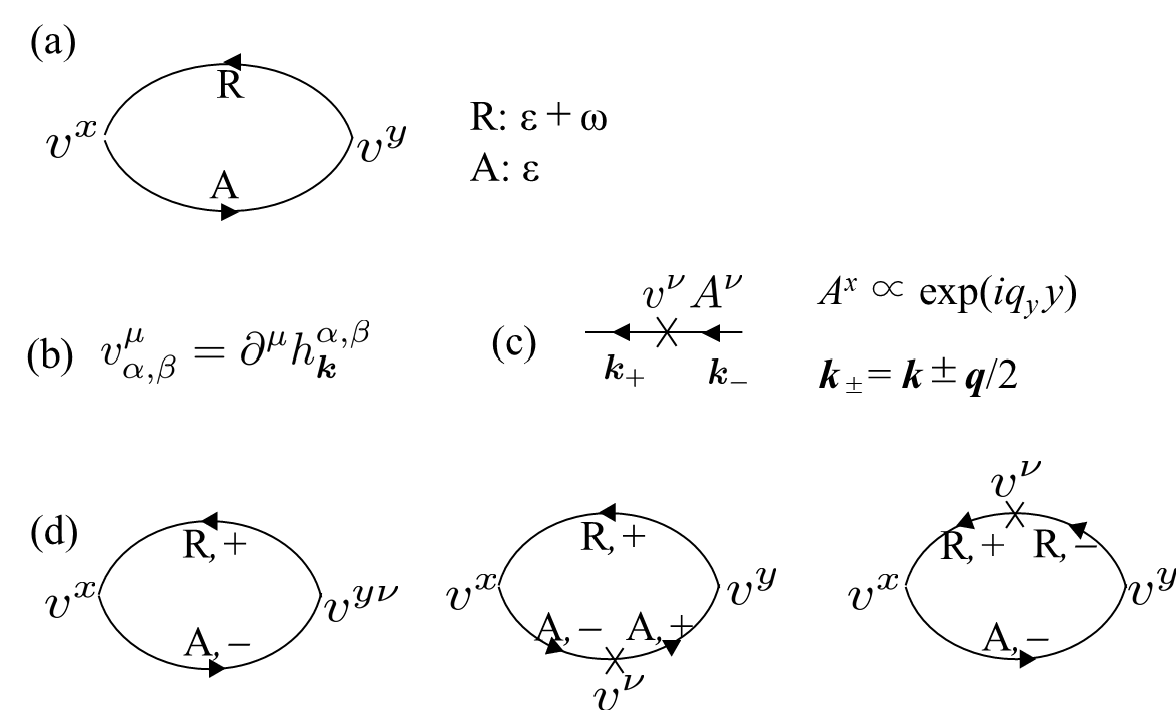}
\caption{
(a) Current correlation function $K(i\w_l)$.
(b) Velocity $v^\mu$ in the orbital representation ($\a,\b$).
(c) $A_\nu$-linear term derived from the Green function, where $\q$ is the momentum of the vector potential and $\k_\pm\equiv\k\pm\q/2$.
(d) $A_\nu$-linear term of (a), composed of two or three Green functions.
}
\label{fig:diagram1}
\end{figure}
%%%%%%%%%%%%%%%%%%%%%%%%%%%%%%%%%%%%%%%

The $xy$-plane Hall conductivity under ${\bm B}\parallel{\bm e}_z$ is defined as
\begin{eqnarray}
j_x=\s_{xy}E_y B_z.
\end{eqnarray}
According to the standard perturbation theory,
\begin{eqnarray}
\s_{xy}=\left.-\frac{\d^2}{i\d\w \d B_z}K(\w)\right|_{\w=B_z=0},
\end{eqnarray}
where $\w$ is the frequency of the electric fields $E_y$.
$K(\w)$ is given by the analytic continuation of the 
current correlation function $K(i\w_l)$:
\begin{eqnarray}
K(i\w_l)&=&g_s T\sum_{n}{\rm Tr}\{ {\hat v}^x {\hat G}^{\rm R}{\hat v}^y {\hat G}^{\rm A} \},
\label{eqn:K-def}
\end{eqnarray}
which is expressed in Fig. \ref{fig:diagram1} (a).
Here, Tr represents the summations of orbital indices and momenta.
${\hat v}$ is the velocity operator and ${\hat G}$ is the Green function for finite ${\bm A}$, R (A) represent the frequency $\e_n+\w_l$ ($\e_n$), and $g_s$ is the spin degeneracy $2$.

To extract the $B_z$-linear term from Eq. (\ref{eqn:K-def}),
we introduce the vector potential ${\bm A}_\q= {\bm A}_0 e^{i{\bm q}\cdot{\bm r}}$,
and take the limit $|\q|\rightarrow0$ at the final stage of the calculation
\cite{Fukuyama,Non-Fermi2}.
Then, the uniform magnetic field is
${\bm B}=[{\rm rot} {\bm A}_\q]_{\q={\bm0}} =[i\q\times{\bm A}_\q]_{\q={\bm0}}$,
or equivalently
$B^\mu= [\e^{\mu\eta\nu}\d^\eta A_\q^\nu]_{\q\rightarrow{\bm0}} = [\e^{\mu\eta\nu}(iq_\eta)A_0^\nu]_{\q\rightarrow{\bm0}}$,
where $|{\bm A}_0|=|{\bm B}|/|\q|$.
For example, 
${\bm B}=(0,0,B_z)$ is obtained when
$\q=(0,q_y,0)$ and ${\bm A}_0=(B_z/iq_y,0,0)$.
Then, the ${\bm A}$-linear potential in the orbital basis is
\begin{eqnarray}
H'&=& e A_0^\nu \sum_{\k} {\hat c}_{\k_-}^\dagger {\hat v}_{\k}^{\nu} {\hat c}_{\k_+}
\label{eqn:Hd}, 
\end{eqnarray}
which are shown in Figs. \ref{fig:diagram1} (b) and (c).
Here, $\k_\pm\equiv
\k\pm\q/2$, and $-e \ (e>0)$ is the carrier of an electron. 
We can also show the following relations up to $O(q^1)$:
\begin{eqnarray}
&&{\hat G}_{\k_\pm}={\hat G}_{\k}\pm\frac12 {\hat G}_{\k}{\hat v}_{\k}^{\mu}{\hat G}_{\k} \cdot q_\mu,
\label{eqn:expand1S}\\
&&{\hat v}_{\k_\pm}^{\mu}= {\hat v}_{\k}^{\mu}\pm\frac12 {\hat {\bar v}}_{\k}^{\mu\nu}\cdot q_\nu,
\label{eqn:expand2S}
%&&{\hat {\tilde v}}_{\k_\pm}^{\a\b}= {\hat {\tilde v}}_{\k}^{\a\b}\pm\frac12 {\hat {\tilde v}}_{\k}^{\a\b\b'}\cdot q_{\b'},
%\label{eqn:expand3S}
\end{eqnarray}
where 
${\hat {\bar v}}_\k^{\mu\nu}\equiv \d^\nu {\hat v}_\k^{\mu}$.
% and ${\hat {\tilde v}}_\k^{\a\b\b'}\equiv \d_{\b'} {\hat {\tilde v}}_\k^{\a\b}$.
%
Thus, the $A$-linear term of Eq. (\ref{eqn:K-def}) is given as
\begin{eqnarray}
K^{(1)}(i\w_l)&=& A_0^\nu g_s T\sum_{\k,n}{\rm tr}\{ {\hat v}^x_\k {\hat
 G}^{\rm R}_{\k+}{\hat {\bar v}}^{y\nu}_{\k} {\hat G}^{\rm A}_{\k-}
\nonumber \\
& &+ {\hat v}^x_\k {\hat G}^{\rm R}_{\k+}{\hat v}^{y}_{\k+} {\hat G}^{\rm A}_{\k+} {\hat v}^\nu_\k{\hat G}^{\rm A}_{\k-} 
\nonumber \\
& &+ {\hat v}^x_\k {\hat G}^{\rm R}_{\k+}{\hat v}^\nu_\k{\hat G}^{\rm R}_{\k-} 
{\hat v}^{y}_{\k-} {\hat G}^{\rm A}_{\k-} \},
\label{eqn:K-def2}
\end{eqnarray}
where tr represents the summation of orbital indices.
This is diagrammatically expressed in Fig. \ref{fig:diagram1} (d).
Here, $\hat{G}_\k\equiv (i\e_n{\hat 1}-{\hat h}_\k)^{-1}$ is the
$M\times M$ electron Green function. 

Next, we derive the expression of $\s_{xy}$ from Eq. (\ref{eqn:K-def2}),
by extracting the $q$-linear term and performing the analytic continuation ($i\w_l\rightarrow \w+i\delta$) \cite{Fukuyama,Non-Fermi2}.
Finally, we convert the obtained expression to the band basis.
In the band basis, the Green function is diagonal:
$G_{a}^{\rm R}(\e)=(\e-\e_{a,\k}+i\gamma_a)^{-1}$.
The velocities in the orbital basis
are converted to those in the band basis ($a,b$) as 
${v}_{ab,\k}^{\mu} \equiv \langle u_{a,\k}|{\hat {v}}_\k^{\mu}|u_{b,\k}\rangle$, and
${\bar v}_{ab,\k}^{\mu\nu}\equiv \langle u_{a,\k}|{\hat v}_\k^{\mu\nu}|u_{b,\k}\rangle$.
We stress that relation ${v}_{a,\k}^{\mu} ={v}_{aa,\k}^{\mu} =\d^\mu \e_{a,\k}$ holds.
%while ${v}_{aa,\k}^{\a\b} \ne \d^\a \d_\b \e_{a,\k}$.

%%%%%%%%%%%%%%%%%%%%%%%%%%%%%%%%%%%%%%%
\begin{figure}[htb]
\includegraphics[width=.99\linewidth]{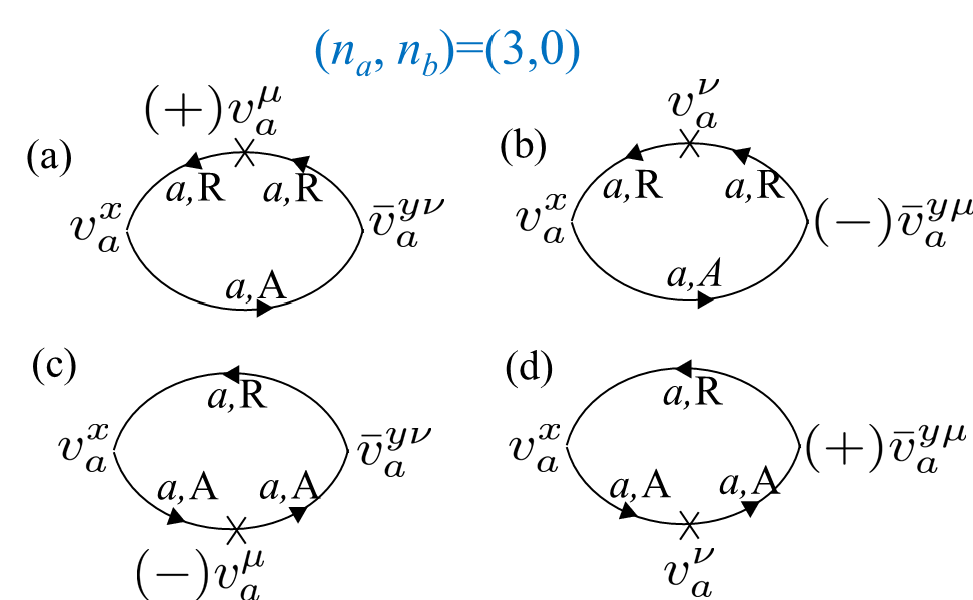}
\caption{
$(n_a,n_b)=(3,0)$ terms for $\s_{xy}$, derived from the $A^\nu q^\mu$-linear terms from $K(i\w_l)$.
$a$ represents the conduction band.
}
\label{fig:diagram2}
\end{figure}
%%%%%%%%%%%%%%%%%%%%%%%%%%%%%%%%%%%%%%%

\begin{figure}[htb]
\includegraphics[width=.99\linewidth]{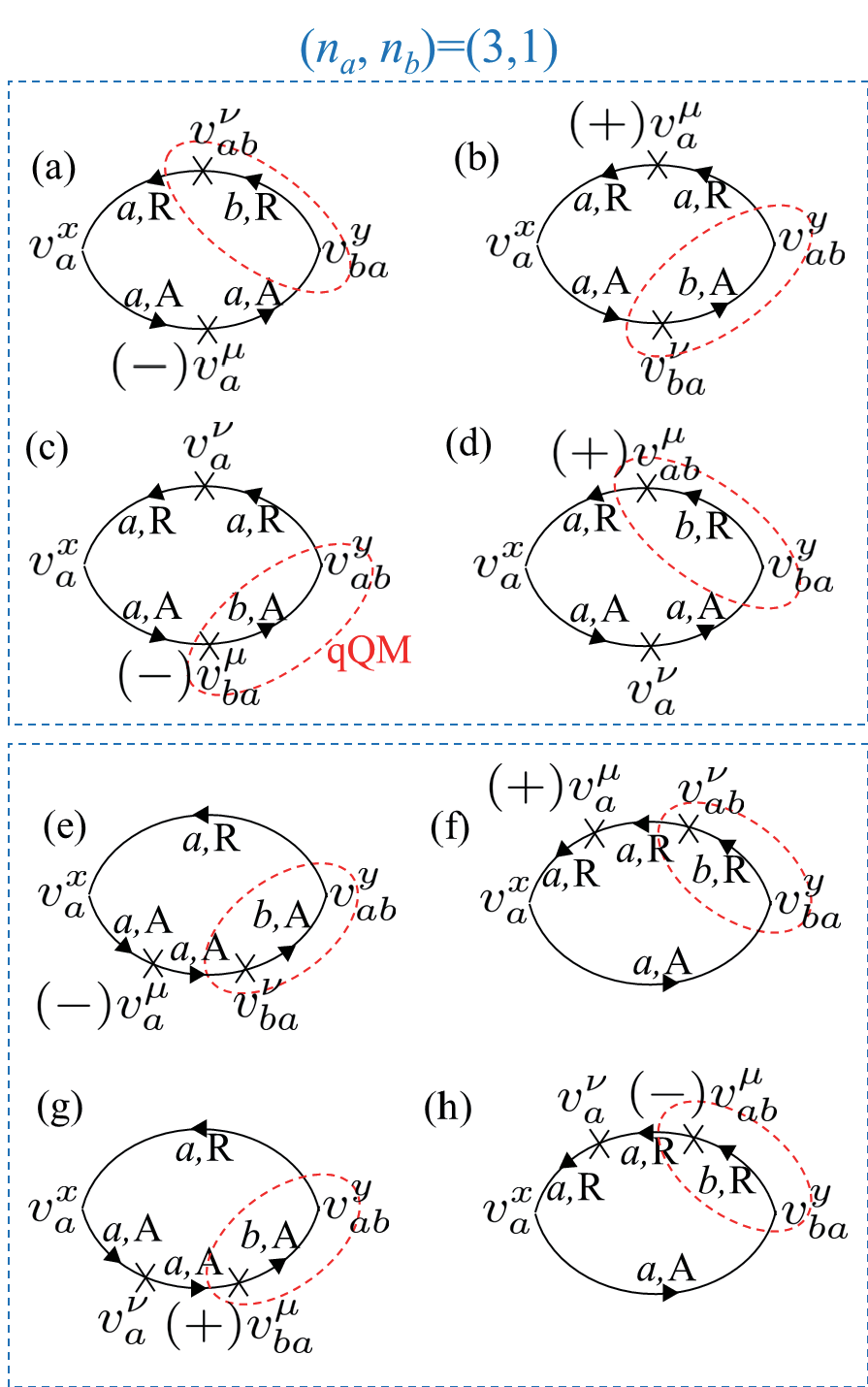}
\caption{
Feynman diagrams of $(n_a,n_b)=(3,1)$ terms for $\s_{xy}$, derived from the $A_\nu q_\mu$-linear terms from $K(i\w_l)$.
 $a$ ($b$) represents the conduction (valence) band.
 The diagrams within the red dashed lines correspond to
 the qQM terms. We stress that (a)+(b)+(c)+(d) is equal to (e)+(f)+(g)+(h).
Other $(n_a,n_b)=(3,1)$ terms are exactly canceled out.
}
\label{fig:diagram3}
\end{figure}

The derived expression of $\s_{xy}$ are composed of $(G_a)^{n_a}\times(G_b)^{n_b}$-terms with $n_a+n_b=3$ or $4$, where $a$ is the conduction band and $b$ is the valence band.
Hereafter, we focus on the $\gamma^{-2}$-term of $\s_{xy}$, which is given by $(n_a,n_b)=(3,0)$ terms shown in Fig. \ref{fig:diagram2} and $(n_a,n_b)=(3,1)$ terms shown in Fig. \ref{fig:diagram3}.
(Note that $(n_a,n_b)=(4,0)$ terms vanish identically, and the terms with $n_a\leq2$ give less divergent contribution.)

Thus, the exact expression for the most divergent $\gamma^{-2}$-term is obtained as
\begin{eqnarray}
\s_{xy}&=& e^3\frac{g_s}{2} \sum_{\k,a}\int\frac{d\e}{2\pi} f'(\e)
{\rm Im}\{(G_a^{\rm R})^2 G_a^{\rm A} - G_a^{\rm R} (G_a^{\rm A})^2\}
\nonumber \\
& &\times v^x_{a}[v^y_a v_a^{xy}(\e) - v^x_a v_a^{yy}(\e)]
\label{eqn:s-xyz} \\
&=& -e^3\frac{g_s}{2} \sum_{\k,a} f'(\e_a) v^x_{a}[v^y_a v_a^{xy}(\e_a) - v^x_a v_a^{yy}(\e_a)]\frac1{2\gamma_a^2},
\label{eqn:s-xyz2}
\end{eqnarray}
where we set $\mu=y$ and $\nu=x$ without loss of generality.
In deriving Eq. (\ref{eqn:s-xyz2}), we used the relation $i\{(G_a^{\rm R})^2
G_a^{\rm A} - G_a^{\rm R} (G_a^{\rm A})^2\} = \frac{\pi}{\gamma_a^2}\delta(\e-\e_a)$
\cite{Fukuyama,Non-Fermi2}.
%The expression (\ref{eqn:s-xyz2}) is unchanged even if the mass-enhancement factor factor $z^{-1}\equiv 1-{\rm Re}\,d\Sigma/\d\e|_{\e=0}$ is taken into account.
Here, the velocity includes the qQM term,
\begin{eqnarray}
v^{\mu\nu}_{a}(\e)&=&{\bar v}^{\mu\nu}_{a}+ 2 g_a^{\mu\nu}(\e).
\label{eqn:vab} 
\end{eqnarray}

%%%%%%%%%%%%%%%%%%%%%%%%%%%%%%%%%%%%%%%
%%%%%%%%%%%%%%%%%%%%%%%%%%%%%%%%%%%%%%%

When $\gamma_b(\e_a\sim0)\ll|\e_a-\e_b|$,
the relation $v^{\mu\nu}_{a,\k}(\e_a)=\d^\mu \d^\nu \e_{a,\k}$ is satisfied,
which is derived from the kinetic equation without interactions.
In this case, Eq. (\ref{eqn:s-xyz2}) reproduces the well-known relaxation-time approximation for single-band models.
In contrast, when $\gamma_b(\e_a\sim0)\gg|\e_a-\e_b|$,
the relation $v^{\mu\nu}_{a,\k}(\e_a)\approx {\bar v}^{\mu\nu}_{a,\k}$
is satisfied.

To summarize, we analyze the normal Hall effect $\s_{xy}$ in strongly correlated multiorbital tight-binding models using the rigorous formula based on the Green-function method.
The formula of $\s_{xy}$ contains the qQM term, which exhibits prominent temperature dependence in strongly correlated metals.
The present qQM mechanism causes the prominent $T$-dependence of $R_H$ in bilayer nickelates.
It is found that $\s_{xy}$ exhibits the coherent-incoherent crossover, from $\s_{xy}^{\rm coh}$ [low-$\rho$ region] to $\s_{xy}^{\rm incoh}$ [high-$\rho$ region].
This is similar to the coherent-incoherent crossover of the anomalous Hall effect: $\s_{\rm AHE}\propto{\rm const}$ for low-$\rho$ region and $\s_{\rm AHE}\propto \rho^{-2}$ for high-$\rho$ region.
\cite{Kontani-AHE}.

In the presence of strong electron correlations,
both the self-energy and the current vertex correction
may play significant roles
\cite{Non-Fermi2,Kontani-AdvPhys}.
As for the self-energy effect, 
the retarded Green function is given as
$G_a^{\rm R}(\k,\e)=(Z_a \e-{\tilde \e}_{a,\k}+i\gamma_a)^{-1}$,
where $Z_a\equiv[(1-{\rm Re}(d\Sigma_a/d\e)_{\e=0}]^{-1}$
is the mass-enhancement factor and 
${\tilde\e}_{a,\k}\equiv \e_{a,\k}+{\rm Re}\Sigma_{a,\k}-\mu$.
Considering the relation $\delta(Z\e-\e_a)=Z^{-1}\delta(\e-\e_a/Z)$,
we can show that $\s_{xy}$ given in Eq. (\ref{eqn:vab}) is independent of the factor $Z_a$.
On the other hand, the current vertex corrections
give rise to a prominent enhancement of the Hall conductivity
and the magneto-conductivity
\cite{Non-Fermi2}.
The effect of the vertex correction for the qQM term
would be a significant future problem.

%%%%%%%%%%%%%%%%%%%%%%%%%%%%%%%%%%%%%%%
\section{Model of thin-film La$_3$Ni$_2$O$_7$} 
%%%%%%%%%%%%%%%%%%%%%%%%%%%%%%%%%%%%%%%
\begin{figure}[!htb]
\includegraphics[width=.99\linewidth]{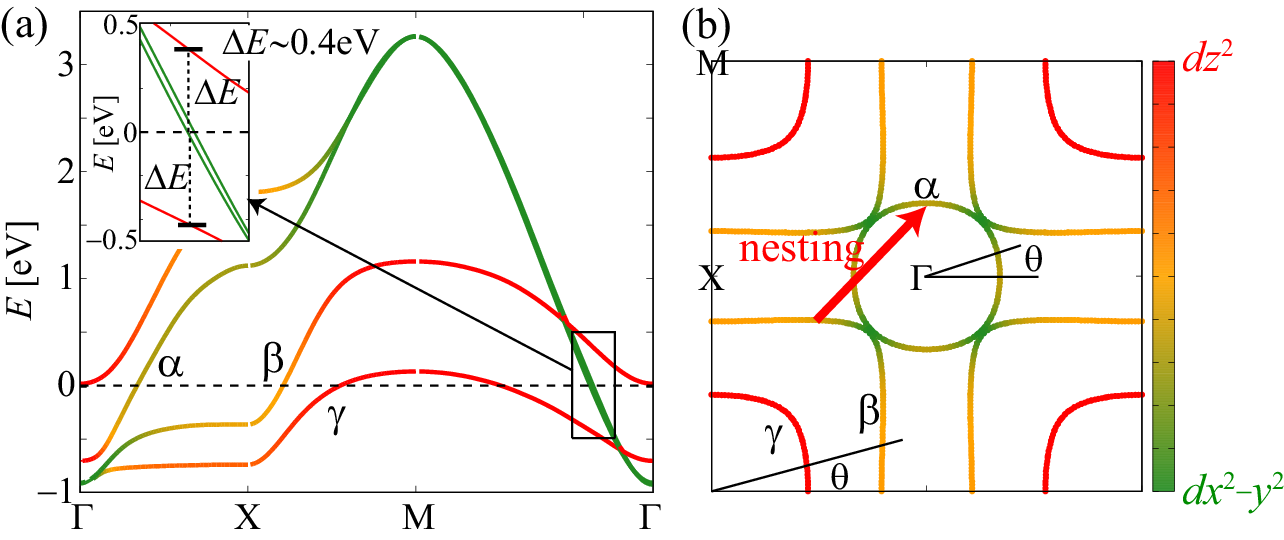}
\caption{
(a) Band structure and
(b) FSs for $n=2.6$ of thin-film $\mathrm{La}_3\mathrm{Ni}_2\mathrm{O}_7$. The color bar indicates the orbital weight
of $d_{z^2}$ and $d_{x^2-y^2}$ orbitals.
The one-electron pocket is labeled by $\alpha$, and the two-hole pockets are labeled by $\beta$ and $\gamma$. $\theta$ denotes the azimuthal angle from the $x$-axis. Red arrows represent the nesting vector.
In the inset of (a), the band energy difference between bands $1(2)$ and $3(4)$ is $\Delta E \sim 0.4$eV. 
} 
\label{fig:fig1}	
\end{figure}
%%%%%%%%%%%%%%%%%%%%%%%%%%%%%%%%%%%%%%

Here, we construct the bilayer 2-orbital tight-binding model 
of thin-film La$_3$Ni$_2$O$_7$. We start with the bulk tight-binding model based on first-principles calculations, as presented in previous work \cite{bilayer-Inoue}.
To reproduce the band
structure near the Fermi energy ($|E|\lesssim 0.5$eV) reported by ARPES measurements in thin-film La$_3$Ni$_2$O$_7$ \cite{Ni-thin-ARPES, Ni-thin-ARPES2}, we reduce the separation between the two $\beta$ FSs facing each other across the X point, while keeping them parallel.
Specifically, we modify
the inter-layer hopping and the level of $d_{z^2}$ orbital, and apply $20$\% hole doping to each Ni atom. Details of the tight-binding model are  explained in Appendix A.
The model Hamiltonian consists of $d_{z^2}$  and $d_{x^2-y^2}$ orbitals
for each upper and lower Ni layer. We denote the $d_{z^2}$ and
$d_{x^2-y^2}$ orbitals in the upper (lower) layer as orbital $l=1(3)$ and $l=2(4)$, respectively.

Figures \ref{fig:fig1} (a) and (b) show the band structure and Fermi surface (FS) in the present thin-film model for electron number $n=2.6$, 
where the color bar indicates the orbital component. 
There are three FSs labeled as $\alpha$, $\beta$, and $\gamma$, where $\alpha$ pocket is an electron pocket around $\Gamma$ point, 
and $\beta$, $\gamma$ pockets are hole pockets around M point.
The $\alpha$ and $\beta$ pockets are the mixture of two $d$ orbitals, while the $\gamma$ pocket is composed of the
$d_{z^2}$ orbital. We also denote the bands including the $\alpha$, $\beta$, and $\gamma$ FSs as the $\alpha$, $\beta$, and $\gamma$ bands, respectively.
Along the $\Gamma$-M line, the two $d_{x^2-y^2}$-orbital bands are nearly degenerate as shown in the
inset of Fig. \ref{fig:fig1} (a). The origin of band degeneracy is the inter-layer hopping of the $d_{x^2-y^2}$ orbital mediated by
$s$-orbital, as previously proposed in YBCO cuprates
\cite{Andersen}. This inter-layer hopping is proportional to $[\cos(k_x)-\cos(k_y)]^2$. Since the inter-layer hopping vanishes along
the $\Gamma$-M line, $d_{x^2-y^2}$-orbital bands on the two layers are nearly decoupled and degenerate.

%%%%%%%%%%%%%%%%%%%%%%%%%%%%%%%%%%%%%%%
\begin{figure}[!htb]
	\includegraphics[width=.99\linewidth]{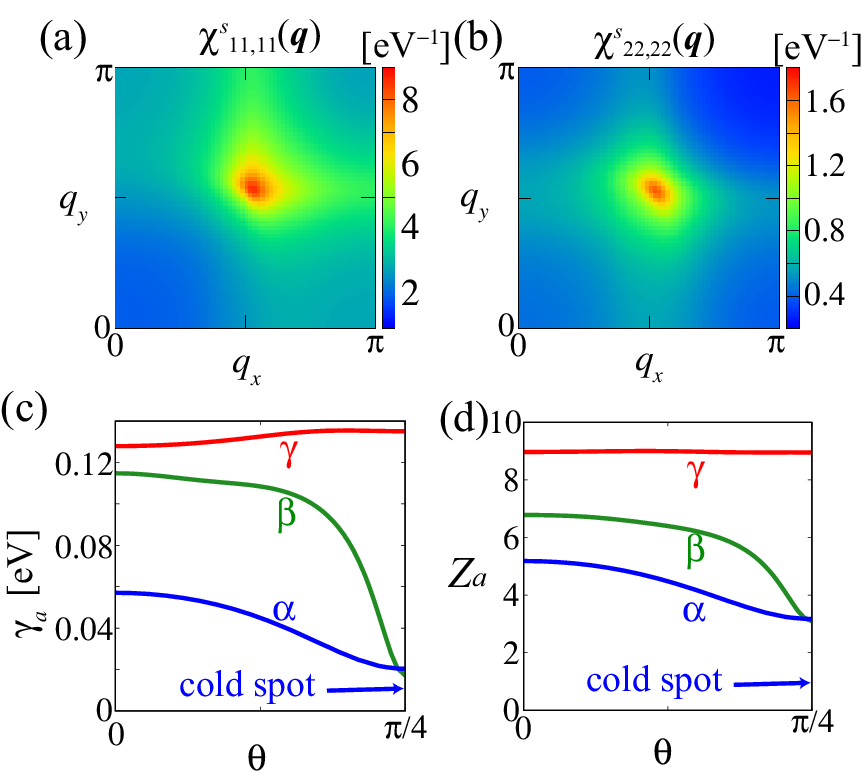}
	\caption{
$\q$ dependences of spin susceptibilities (a) $\chi^s_{11,11}(\q)$ and (b) $\chi^s_{22,22}(\q)$
 for $U=2.74$eV, $J/U=0.2$ ($\alpha_s=0.97$ at
 $T=10$meV) in
 the FLEX approximation.
 $\theta$ dependences of (c) damping $\gamma_a$ and (d) mass enhancement
 $Z_a$ on the FSs in the band
 representation for $U=2.74$eV, $J/U=0.2$ at $T=10$meV.
}
\label{fig:fig2}
\end{figure}
%%%%%%%%%%%%%%%%%%%%%%%%%%%%%%%%%%%%%%%

%%%%%%%%%%%%%%%%%%%%%%%%%%%%%%%%%%%%%%%
\section{The results of FLEX and resistivity}
%%%%%%%%%%%%%%%%%%%%%%%%%%%%%%%%%%%%%%%
Here, we analyze the self-energy $\hat{\Sigma}$ in the fluctuation-exchange (FLEX)
approximation. Details of formulation are explained in Appendix B. 
In the FLEX approximation, the spin susceptibilities are dominant over the charge susceptibilities.
Therefore, the FLEX self-energy mainly originates from spin susceptibilities. Thus, the spin susceptibilities are important to understand the nature of the quasiparticle damping and cold (hot) spots.
Figures \ref{fig:fig2} (a) and (b) show the obtained intra-layer spin
susceptibilities $\chi^s_{11,11}(\bm{q},0)$ and
$\chi^s_{22,22}(\bm{q},0)$ for $U=2.74$eV, $J/U=0.2$ at $T=10$meV in the FLEX approximation, where the Stoner factor is $\a_s=0.97$.
$\chi^s_{11,11}(=\chi^s_{33,33})$ is dominant over
$\chi^s_{22,22}(=\chi^s_{44,44})$ and takes large values at
$\bm{q}\sim(\pi/2,\pi/2)$. This peak position corresponds to the
nesting vector shown in Fig. \ref{fig:fig1} (b).
The inter-layer susceptibility $\chi^s_{11,33}$ (not shown) is negative and a
little smaller than $\chi^{s}_{11,11}$.
Similar results have been reported in bulk La$_3$Ni$_2$O$_7$ \cite{Lechermann-Ni-SC, Luo-Ni-SDW, Eremin-Ni-SDW}.

\begin{figure}[!htb]
	\includegraphics[width=.99\linewidth]{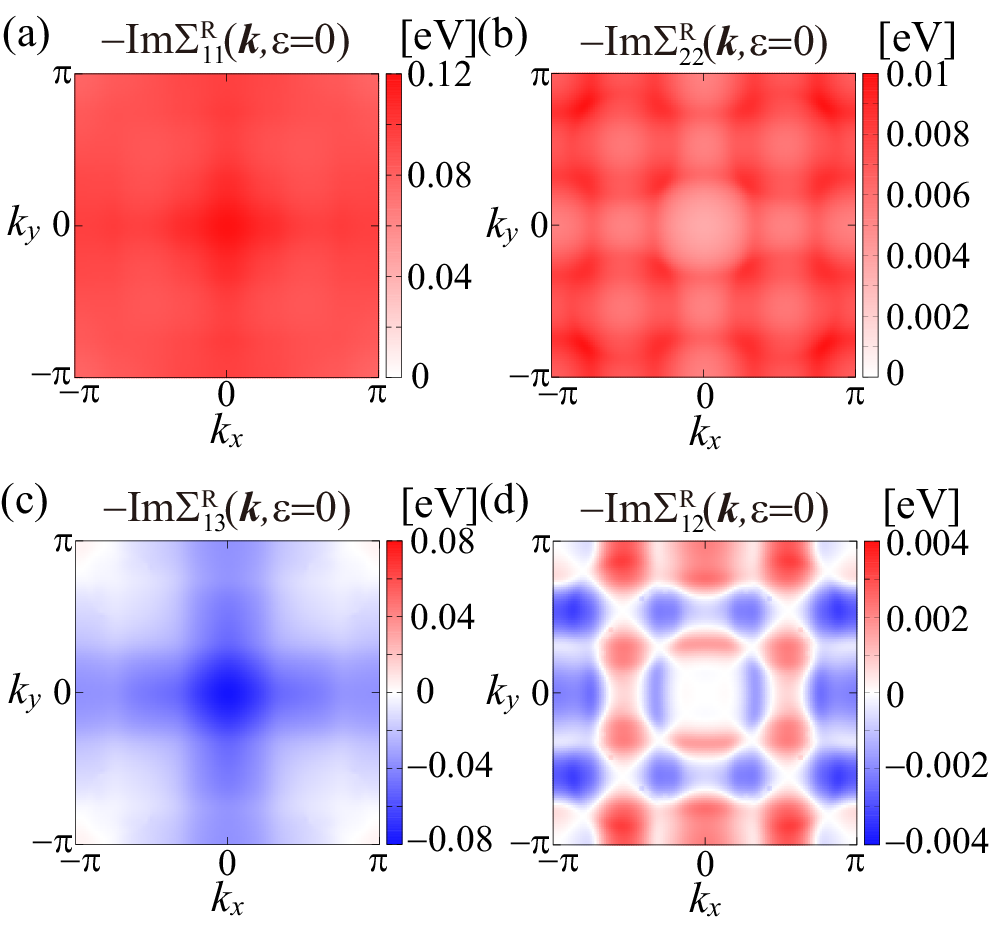}
	\caption{$\bm{k}$ dependences of $-{\rm Im}\Sigma^{\rm R}_{lm}(\bm{k},\epsilon=0)$ in the orbital representation for (a) $(l,m)=(1,1)$, (b) $(l,m)=(2,2)$, (c) $(l,m)=(1,3)$, and (d) $(l,m)=(1,2)$.  
}
\label{fig:fig15}
\end{figure}

Figures \ref{fig:fig2} (c) and (d) show the obtained $\theta$ dependence of
damping $\gamma_a\equiv -{\rm
Im}\Sigma^{\rm R}_a$ and mass-enhancement factor $Z_a\equiv 1-{\rm
Re}\frac{\d\Sigma^{\rm R}_a(\e)}{\d\e}|_{\e=0}$ on the FSs in the band representation for $U=2.74$eV,
$J/U=0.2$ at $T=10$meV, $\theta$ denotes the azimuthal angle from the $x$-axis.
The behaviors of $\gamma_a$ and $Z_a$ are similar.
The $\alpha$ and $\beta$ FSs become the cold spots at $\theta=\pi/4$,
where the weight of the $d_{x^2-y^2}$ orbital is dominant.
As shown in Fig. \ref{fig:fig15}, the self-energy for $d_{x^2-y^2}$ orbital is much smaller than that for $d_{z^2}$ orbital, since $\chi^s_{22,22}$ is much smaller than $\chi^s_{11,11}$.
Thus, the $d_{x^2-y^2}$-orbital cold spots emerge in La$_3$Ni$_2$O$_7$.
We also note that $\bm{k}$ dependence of  $-{\rm Im}\Sigma^{\rm R}_{11}(\bm{k},\epsilon=0)$ is weak.

%%%%%%%%%%%%%%%%%%%%%%%%%%%%%%%%%%%%%%%
%%%%%%%%%%%%%%%%%%%%%%%%%%%%%%%%%%%%%%

At the cold spot, $b$-band states, where the smaller value of $b$ corresponds to the lower-energy band, are expressed in the $l$-orbital basis as $|b=1\rangle=\frac{|l=1\rangle+|l=3\rangle}{\sqrt{2}}$, 
$|b=2\rangle=\frac{|l=2\rangle-|l=4\rangle}{\sqrt{2}}$, 
$|b=3\rangle=\frac{|l=2\rangle+|l=4\rangle}{\sqrt{2}}$, and
$|b=4\rangle=\frac{|l=1\rangle-|l=3\rangle}{\sqrt{2}}$.
We note that $|b=1\rangle$ and $|b=3\rangle$ possess even parity under two-layer inversion, while  $|b=2\rangle$ and $|b=4\rangle$ possess odd parity.

Using the self-energy in the FLEX approximation, transport phenomena are analyzed by the linear response theory.
 We calculate $\rho$, $R_H$, Nernst
 coefficient $\nu$, and the Seebeck coefficient $S$ in
 the band representation.
 Details of the formulation are explained in Appendix B.
 Figure \ref{fig:figS4} (a) exhibits $T$-dependences of
$\rho$ for $J/U=0.25$, $0.2$, and $0.15$, where $\alpha_s=0.95$ is
satisfied at $T=10$meV.
Non-Fermi-liquid behavior $\rho\propto T$ is obtained due to the strong
spin fluctuations. The origin of the $T$ dependence is $\gamma\propto T$
 around the cold spots as
 shown in Fig. \ref{fig:figS4} (b). This
result is consistent with the experimental results \cite{Ko-Ni-SC, Ni-thin-trans}.
Non-Fermi-liquid behavior $\rho\propto T$ has also been obtained in other strongly
correlated electron systems.
 
%%%%%%%%%%%%%%%%%%%%%%%%%%%%%%%%%%%%%%%
\begin{figure}[!htb]
	\includegraphics[width=.99\linewidth]{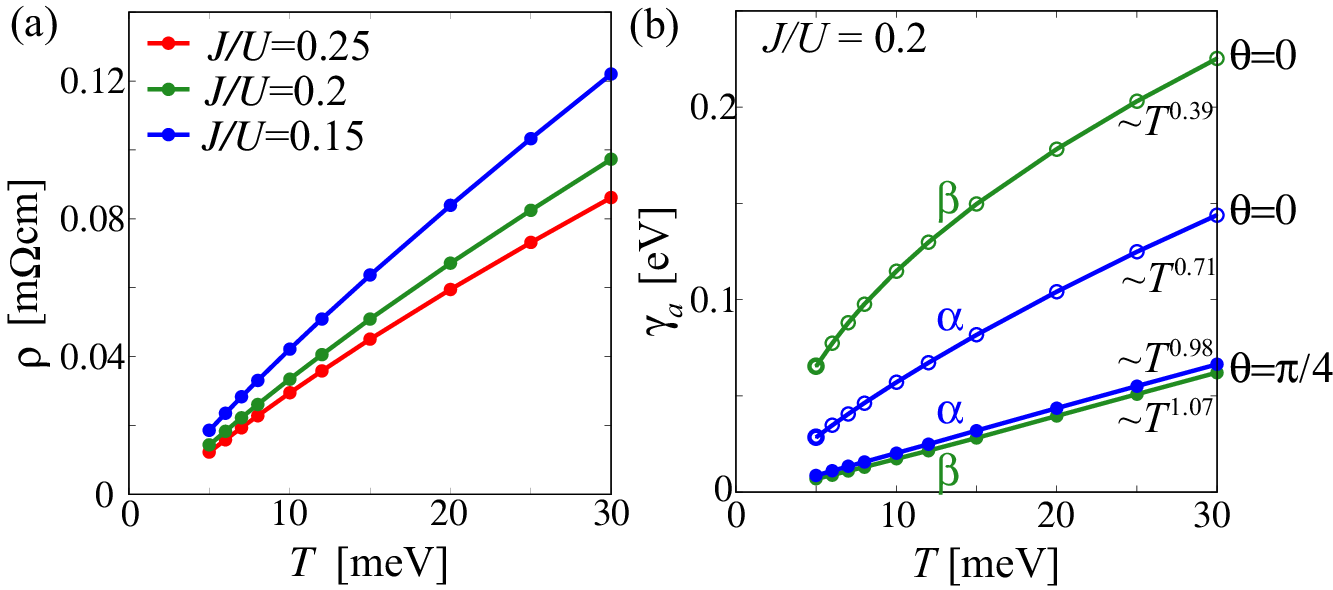}
 \caption{
 (a) $T$ dependences of $\rho$ for $J/U=0.25, 0.2, 0.15$, where
 $\a_s=0.95$ is satisfied at $T=10$meV. 
(b) $T$ dependences of $\gamma_a$ for $\theta=0,\pi/4$ on $\alpha$ and
 $\beta$ FSs for $U=2.74$eV, $J/U=0.2$.
}
\label{fig:figS4}
\end{figure}
%%%%%%%%%%%%%%%%%%%%%%%%%%%%%%%%%%%%%%%

\section{Results of Hall coefficient and Nernst coefficient}
Here, we analyze the transport coefficient using the formula that incorporates the qQM term $g^{\mu\nu}_a[\gamma]$. This qQM term, which fully
captures many-body effects by accounting for the $T$ dependence of
$\gamma$, plays a crucial role in determining $R_H$ and $\nu$.
$g^{\mu\nu}_a$ term in
Eq. (\ref{eqn:qqm}) is
composed of the interband terms such as $v^\mu_{ab}$. As shown in
Sec. II, the result of the
conventional RTA is reproduced by using
$g^{\mu\nu}_a[0]$. On the other hand, the result given by using $g^{\mu\nu}_a[\infty]=0$
corresponds to the single-band approximation, where the qQM term is absent.
In the following, we denote the Hall coefficients obtained by
 using $g^{\mu\nu}_a[\gamma]$, $g^{\mu\nu}_a[0]$, and
 $g^{\mu\nu}_a[\infty]$ as $R_H$, $R_H^{\rm RTA}$, and $R_H^{\rm no \, qQM}$, respectively.
We also denote the Nernst coefficients obtained by
 using $g^{\mu\nu}_a[\gamma]$, $g^{\mu\nu}_a[0]$, and
 $g^{\mu\nu}_a[\infty]$ as $\nu$, $\nu^{\rm RTA}$, and $\nu^{\rm no \, qQM}$, respectively.
 
\begin{figure}[!htb]
	\includegraphics[width=.99\linewidth]{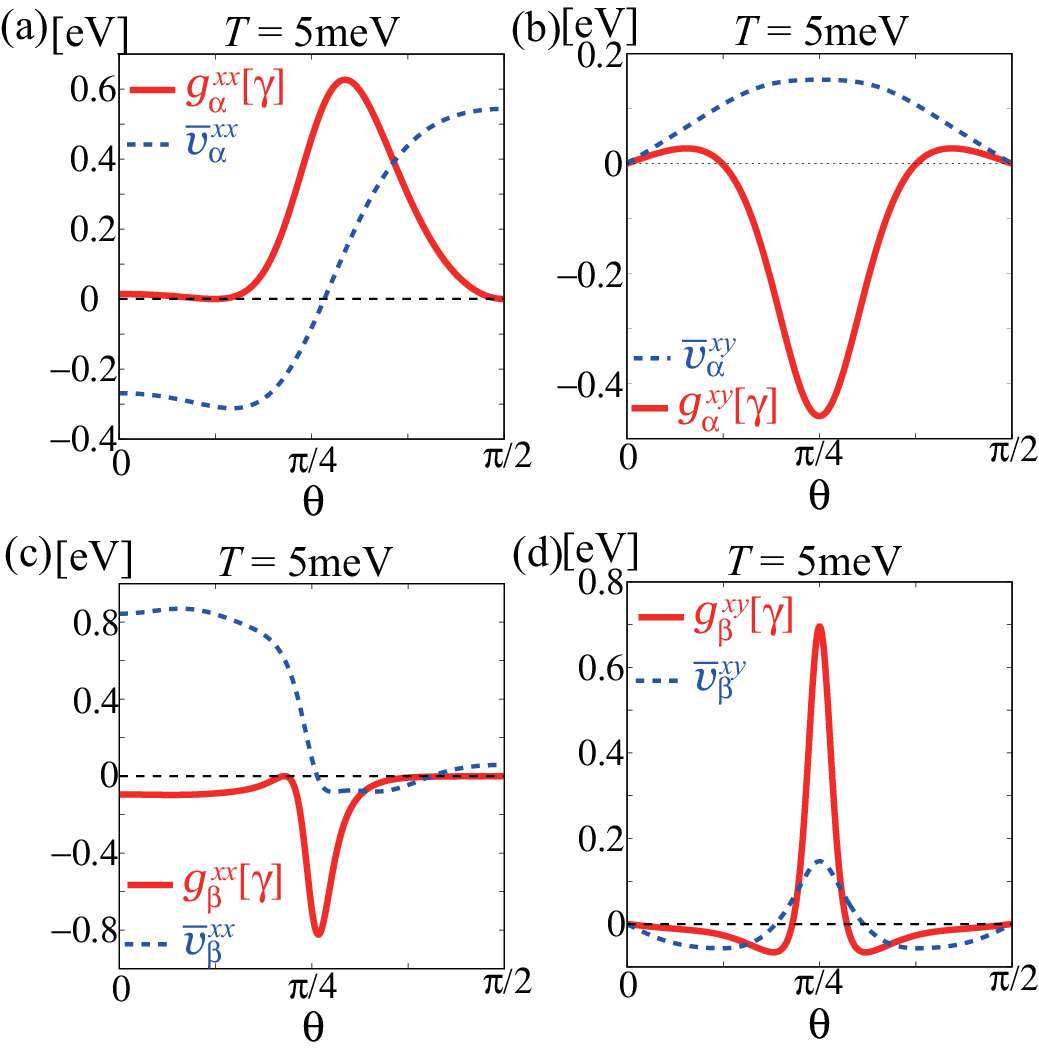}
	\caption{
 $\theta$ dependences of (a) $g_\a^{xx}[\gamma]$ and
 $\bar{v}_\a^{xx}$, (b) $g_\a^{xy}[\gamma]$ and
 $\bar{v}_\a^{xy}$, (c) $g_\b^{xx}[\gamma]$ and
 $\bar{v}_\b^{xx}$, and (d) $g_\b^{xy}[\gamma]$ and
 $\bar{v}_\b^{xy}$ on FSs $(\e=\mu)$ for $U=2.74$eV, $J/U=0.2$ at $T=5$meV. $\theta=\pi/4$ corresponds
 to the cold spot.
}
\label{fig:v-g}
\end{figure}

Figure \ref{fig:v-g} shows $\theta$ dependences of
$g^{\mu\nu}_a[\gamma]$ and $\bar{v}^{\mu\nu}_a$ on
FSs $(\e=\mu)$ for
$U=2.74$eV, $J/U=0.2$
 at $T=5$meV. The value of $g^{\mu\nu}_a[\gamma]$ at cold spot
 $(\theta=\pi/4)$ is significantly
 larger than that of $\bar{v}^{\mu\nu}_a$ at low temperatures.
 This behavior is understood by the expression for Re$G$ in
 Eq. (\ref{eqn:ReG}), which is enhanced when $\gamma_b$ and
 $\mu-\e_b$ are small.
 In La$_3$Ni$_2$O$_7$, at cold spots, $g^{\mu\nu}_{a=2}$ $(\mu,\nu=x,y)$ in $\beta$ band is induced by $b=4$ band, since these bands possess the same odd parity under two-layer inversion. In contrast, $g^{\mu\nu}_{a=3}$ in $\alpha$ band is induced by $b=1$ band, which has the same even parity.  Thus, $g^{\mu\nu}_{a}$ for inter-degenerate bands with different parities are absent for $\mu,\nu=x,y$. However, we note that $g^{xz}_a$ and $g^{zz}_a$ remain finite for the inter-degenerate bands, which leads to a finite $\sigma_{xz}$.
As shown in the inset of Fig. \ref{fig:fig1}, the relation $\Delta E=\mu-\e_b\sim 0.4$eV between bands with the same parity is obtained at cold spots. At low temperatures, small value of $\gamma_b(\e=0)(<\Delta E)$ enhances $g^{\mu\nu}_{a}$.
In contrast, at high temperatures, the magnitude of $g^{\mu\nu}_a[\gamma]$ becomes smaller than its low-temperature value ($\sim g^{\mu\nu}_a[0]$), as shown in Fig. \ref{fig:fig13} in Appendix C. This behavior arises because $\gamma_b(\e=0)$ reaches $\Delta E$ at high temperatures, even for the large value of $\Delta E\sim 0.4$eV.

%%%%%%%%%%%%%%%%%%%%%%%%%%%%%%%%%%%%%%%
%%%%%%%%%%%%%%%%%%%%%%%%%%%%%%%%%%%%%%

\begin{figure}[!htb]
	\includegraphics[width=.99\linewidth]{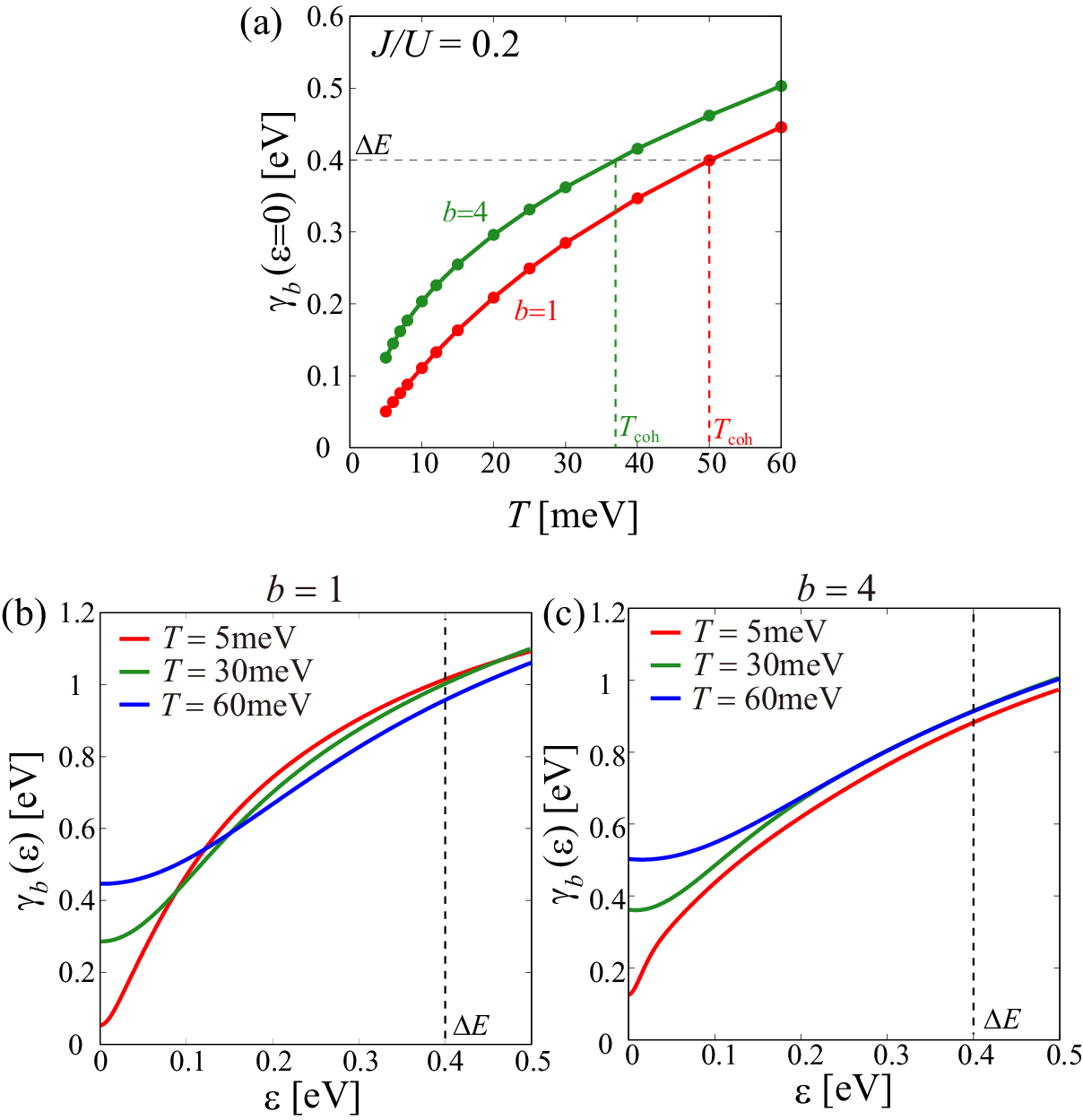}
	\caption{
(a) $T$ dependences of $\gamma_b(\e=0)$ at cold spot for band $b=1,4$ for $U=2.74$eV, $J/U=0.2$.
$T_{\rm coh}\sim 50$meV for $b=1$ and $T_{\rm coh}\sim 37$meV for $b=4$ are estimated by $\Delta E\sim 0.4$eV. 
(b) $\e$ dependences of $\gamma_b(\e)$ at cold spot for band $b=1$ at $T=5,30,60$meV, and (c) those for $b=4$.
}
\label{fig:fig14}
\end{figure}

Figure \ref{fig:fig14} shows $T$ dependences of $\gamma_b(\e=0)$ at the cold spot for bands $b=1$ and $4$ for $U=2.74$eV, $J/U=0.2$.
The values of $\gamma_b(\e=0)$ for $b=1,4$ are larger than those for $b=2,3$ ($\beta$ and $\alpha$ bands), as shown in Fig. \ref{fig:figS4} (b), since $b=1,4$ bands at the cold spot are mainly composed of $d_{z^2}$ orbitals.
Here, $T_{\rm coh}$ is estimated from the condition $\gamma_b(\e=0)\sim \Delta E\sim 0.4$eV. We obtain $T_{\rm coh}\sim 50(37)$meV for $b=1(4)$. The strong $T$ dependence of $\gamma_b(\e=0)$ for $b=1,4$ originates from $\e=0$, as shown in Figs. \ref{fig:fig14} (b) and (c). We confirm that the $T$ dependence of $\gamma_b(\e)$ is small for $\e\sim \Delta E$, where the $b=1,4$ bands are located away from the Fermi energy.

\begin{figure}[!htb]
	\includegraphics[width=.99\linewidth]{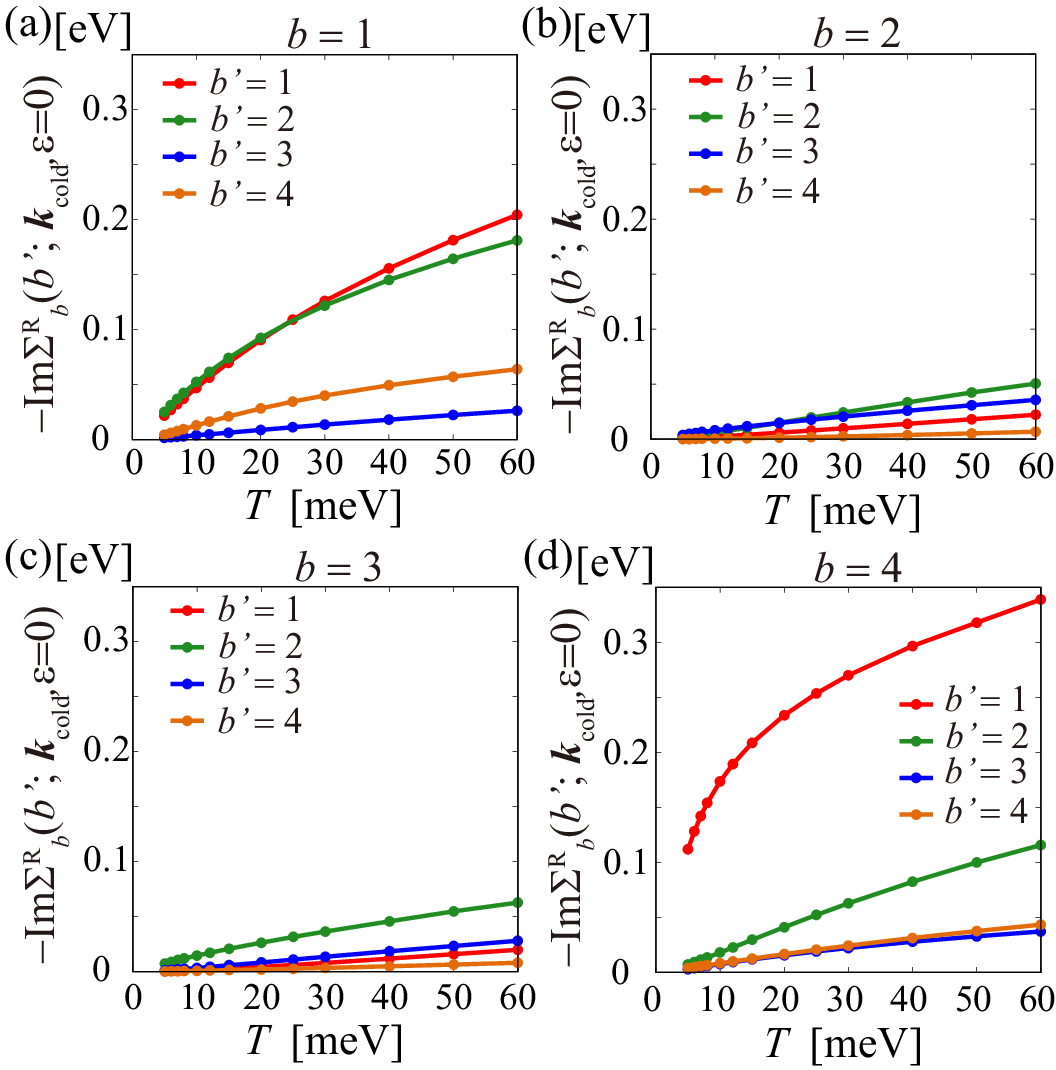}
	\caption{
$T$ dependences of $-{\rm Im}\Sigma^{\rm R}_b(b';k_{\rm cold},\e=0)$ for (a) $b=1$, (b) $b=2$, (c) $b=3$, and (d) $b=4$. 
}
\label{fig:fig16}
\end{figure}

To understand the details of $\gamma_b(\e=0)$ at the cold spot for $b=1,4$, we analyze the band dependence of the interband FLEX interaction contributing to the self-energy. Using the analytic continuation, the contribution to the self-energy of band $b$ at the cold spot from band $b'$ is given by
\begin{flalign}
&-{\rm Im}\Sigma_b^{\rm R}(b';\bm{k}_{\rm cold},\epsilon=0)=\nonumber\\ 
&\frac{1}{(2\pi)^2}\int_{k'_{FS-b'}} \frac{d k'_{FS-b'}}{|v_{k'_{FS-b'}}|}\int \frac{d \epsilon'}{2}  
\left[\tanh\frac{\epsilon'}{2T}-{\rm cotanh}\frac{\epsilon'}{2T}\right]\nonumber\\
&\hspace{3cm}\times {\rm Im}V^{\rm FLEX}_{b,b'}(\bm{k}_{\rm cold},\bm{k}',-\epsilon'),
\end{flalign}
where $V^{\rm FLEX}_{b,b'}$ denotes the interband interaction between bands $b$ and $b'$ in the FLEX approximation, and $k_{FS-b}$ denotes momentum on the band-$b$ FS. From this equation, $T$-dependence of $-{\rm Im}\Sigma^{\rm R}_b(b';\bm{k}_{\rm cold},\epsilon=0)$ originates from $T^2(\partial/\partial\e') {\rm Im}V^{\rm FLEX}_{b,b'}(\bm{k}_{\rm cold},\bm{k}',-\epsilon')|_{\e'=0}\propto T$. 
As shown in Fig. \ref{fig:fig16}, for $b=1$, $-{\rm Im}\Sigma^{\rm R}_b(b';\bm{k}_{\rm cold},\epsilon=0)$ is dominated by the contributions from $b'=1,2$, whereas for  $b=4$ it is dominated by the contribution from $b'=1$. Thus, the large value and $T$-dependence of $\gamma_b(\e=0)$ mainly originate from scattering from the $\gamma$ pocket. We also note that the $\bm{k}$-independent $\gamma_{11}$ in the orbital representation shown in Fig. \ref{fig:fig15} (a) is another factor contributing to the large $\gamma_b(\e=0)$ for $b=1,4$ at the cold spot.

%%%%%%%%%%%%%%%%%%%%%%%%%%%%%%%%%%%%%%%

%%%%%%%%%%%%%%%%%%%%%%%%%%%%%%%%%%%%%%%
%%%%%%%%%%%%%%%%%%%%%%%%%%%%%%%%%%%%%%

\begin{figure}[!htb]
	\includegraphics[width=.99\linewidth]{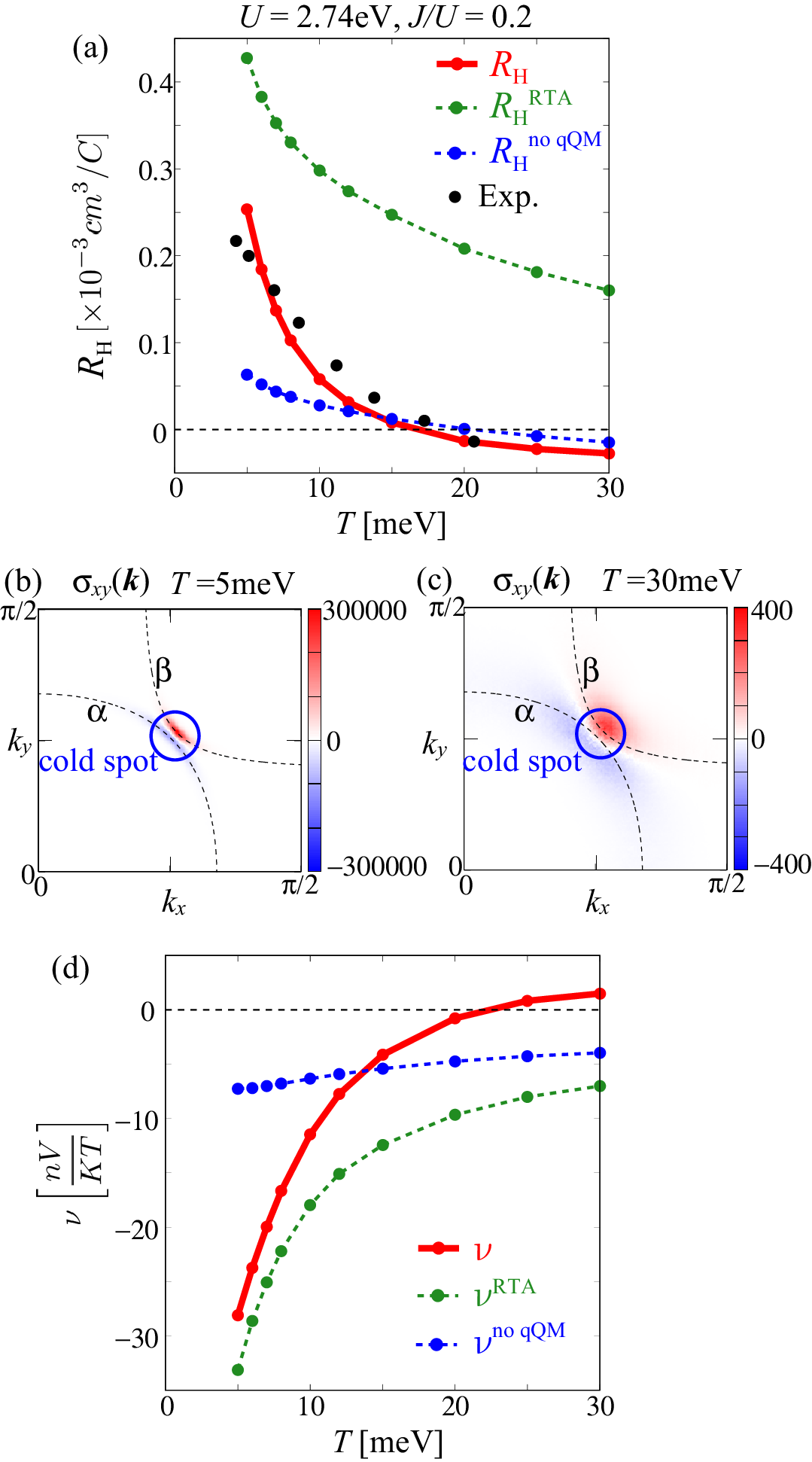}
	\caption{
(a) $T$ dependences of  $R_H$, $R_H^{\rm RTA}$, and $R_H^{\rm no \, qQM}$ for $U=2.74$eV, $J/U=0.2$. The experimental
 results reported in Ref. \cite{Ko-Ni-SC} are plotted as black dots.
 $\k$ dependences of $\sigma_{xy}(\k)$ obtained by using
 $g^{\mu\nu}_a[\gamma]$ at (b) $T=5$meV and (c) $T=30$me for
 $U=2.74$eV, $J/U=0.2$. Dashed lines represent FSs. Blue circles
 represent cold spots.
 (d) $T$ dependences of  $\nu$, $\nu^{\rm RTA}$, and $\nu^{\rm no \, qQM}$ for $U=2.74$eV, $J/U=0.2$.
}
\label{fig:fig3}
\end{figure}

Figure \ref{fig:fig3} (a) shows $T$ dependences of $R_H$, $R_H^
{\rm RTA}$, and $R_H^{\rm no \, qQM}$ for $U=2.74$eV, $J/U=0.2$.
  The experimental
 results reported in Ref. \cite{Ko-Ni-SC} are also included for
 comparison.
 The values of $R_H$ and $R_H^{\rm RTA}$ at low $T$ are much larger than that of $R_H^{\rm no \, qQM}$, which means that the qQM terms are important for $R_H$, including the second derivative velocity $v^{\mu\nu}$.
The condition $\gamma_b \sim \Delta E\gg T$, arising from the strong temperature dependence of $\gamma_b$, is essential for a pronounced temperature dependence of $R_H$, which exhibits a stronger temperature dependence than $R_H^{\rm RTA}$.
As shown in Fig. \ref{fig:fig14} (a), the relation $\gamma_b \sim \Delta E$ ($\sim 0.4$eV)
 is satisfied at $T=T_{\rm coh}\sim 50(37)$meV for $b=1(4)$.
For $T\ll T_{\rm coh}$, $R_H$ is close to $R_H^{\rm
RTA}$ due to $\gamma_b<\Delta E$.
In contrast, for $T\gtrsim T_{\rm coh}$, $\gamma_b$ becomes large $(\gamma_b>\Delta E)$,
 and $R_H$ approaches $R_H^{\rm no \, qQM}$.
We note that the decrease in $R_H$ at temperatures much lower than $T_{\rm coh}$ originates from the fact that $T_{\rm coh}$ for $b = 4$ is lower than that for $b = 1$. As a result, the positive contribution to $R_H$ from $a=2$ ($\beta$ band) is more strongly suppressed by $\gamma_{b=4}$ than the negative contribution from $a=3$ ($\alpha$ band), which is suppressed by $\gamma_{b=1}$.
This reflects the strong competition between $\alpha$ and $\beta$ bands, as shown later in Fig. \ref{fig:fig5} (a).
Consequently, $R_H$ exhibits a crossover from $R_H^{\rm RTA}$ with small $\gamma_b$ in the low-$T$ coherent regime to $R_H^{\rm no \, qQM}$ with large $\gamma_b$ in the high-$T$ incoherent regime.
Due to this coherent-incoherent crossover behavior, discussed in
Sec. II, the $T$ dependence of $R_H$ is more significant than that of
$R_H^{\rm RTA}$.
Notably, the obtained $T$ dependence of $R_H$ is consistent with experimental results in thin-film bilayer nickelates \cite{Ko-Ni-SC, Ni-thin-trans}.

Figures \ref{fig:fig3} (b) and (c) show $\k$ dependences
of $\sigma_{xy}(\k)$, where $\sigma_{xy}$ is given by $\sigma_{xy}=\frac{1}{N}\sum_{\k}\sigma_{xy}(\k)$.
These results indicate that the dominant contributions to $R_H$ originate from the cold spots.
 The hole pocket
$\beta$ contributes positively to $R_H$, whereas the electron pocket
$\alpha$ contributes negatively to $R_H$. 
Competition between the contributions from $\alpha$ and $\beta$ pockets
is important to determine both the sign and $T$ dependence of $R_H$. At low
$T$, the contribution from the vicinity of the FS at cold spots is
dominant.  Furthermore, $\gamma$ on hole pocket
$\beta$ at cold spots
is smaller than that on electron pocket $\alpha$, as shown in Fig. \ref{fig:fig2} (c).
In this case, the positive contribution is dominant over the negative one, as shown in
Fig. \ref{fig:fig3} (b). In contrast, at high $T$, the contributions
from the two pockets are comparable, since 
the negative contribution from the $\alpha$ pocket relatively develops away from the cold spots, as shown in
Fig. \ref{fig:fig3} (c).

Figure \ref{fig:fig3} (d) exhibits $T$ dependences of $\nu$, $\nu^{\rm
RTA}$, and $\nu^{\rm no \, qQM}$ for $U=2.74$eV, $J/U=0.2$.
At low T, $\nu$ and $\nu^{\rm RTA}$ are significantly
larger in magnitude than $\nu^{\rm no \, qQM}$, which also means that the qQM terms are
important for $\nu$ including the second derivative velocity
$v^{\mu\nu}$.
In contrast to $R_H$, $\nu$ is negative, which decreases at low $T$. 
%\textcolor{blue}{$\nu$ is close to $\nu^{\rm RTA}$ for $T<T_{\rm coh}$, while  $\nu$ is close to $\nu^{\rm no \, qQM}$ for $T>T_{\rm coh}$. 
The coherent-incoherent crossover behavior in $\nu$ is similar to that of $R_H$. 
\color{black}

%%%%%%%%%%%%%%%%%%%%%%%%%%%%%%%%%%%%%%%
%%%%%%%%%%%%%%%%%%%%%%%%%%%%%%%%%%%%%%

\section{Discussion}
\begin{figure}[!htb]
	\includegraphics[width=.99\linewidth]{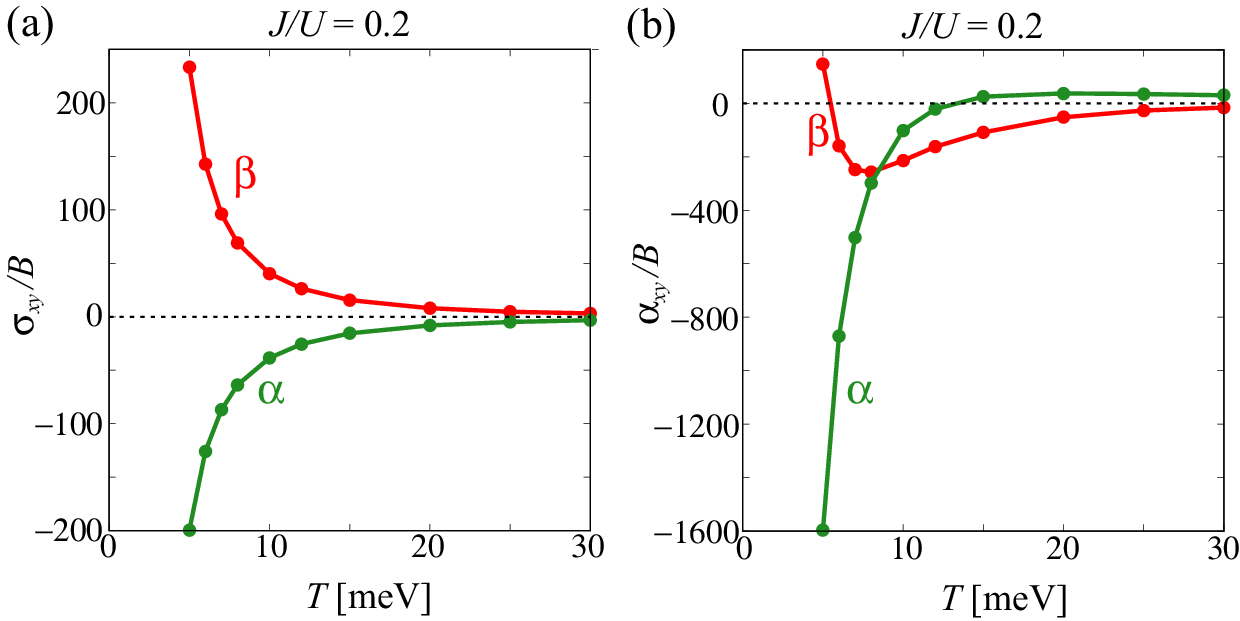}
	\caption{
 $T$ dependences of (a) $\sigma_{xy}/B$ and (b) $\alpha_{xy}/B$ for
 $\alpha$ and $\beta$ bands obtained by using
 $g^{\mu\nu}_a[\gamma]$ for $U=2.74$eV with $J/U=0.2$.
}
\label{fig:fig5}
\end{figure}

Here, we discuss the competing contributions to transport phenomena from
the $\alpha$ and the $\beta$ bands. Figures \ref{fig:fig5} (a) and (b) show the contributions to $\sigma_{xy}/B$ and
$\alpha_{xy}/B$ from $\alpha$ and $\beta$ bands obtained by using
 $g^{\mu\nu}_a[\gamma]$ for $U=2.74$eV, $J/U=0.2$.
We find that the positive contribution from the $\beta$ band to $\sigma_{xy}$ strongly competes with the negative contribution from the $\alpha$ band. 
As a result, the magnitude of $R_H$ in La$_3$Ni$_2$O$_7$ is much smaller than that in cuprates with only a hole band.
Due to the strong competition, $T$ dependence of $R_H$ becomes significant. Consequently, the sign of $R_H$ is sensitive to model parameters such as $U$ and $J$, as shown in Fig. \ref{fig:fig4}(a) in Appendix C.
In contrast, the almost perfect cancellation of the $\a$ and $\b$ bands in
$\sigma_{xy}$ does not exist in $\a_{xy}$. Therefore, negative $\a_{xy}$
and $\nu$,
mainly given by $\alpha$ band, are
greatly enhanced at low temperatures. These behaviors are robust against
the value of $U$ as shown in Fig. \ref{fig:fig5-2} in Appendix C. The possible reasons for
incomplete cancellation in $\a_{xy}$ are (i) asymmetry with respect to
energy $\e$ in $\gamma_a$ and (ii) contributions away from the cold spots are
different between $\alpha$ and $\beta$ bands.

%%%%%%%%%%%%%%%%%%%%%%%%%%%%%%%%%%%%%%%
%%%%%%%%%%%%%%%%%%%%%%%%%%%%%%%%%%%%%%

We also calculate Seebeck coefficient $S$ as
shown in Fig. \ref{fig:figS6} in Appendix C.
The obtained $S$ is negative and increases with decreasing $T$, which is
similar to the behavior observed in triple-layer and quintuple-layer
nickelates \cite{Ni-S}.

Here, we note that the transport phenomena are not affected qualitatively
by the CDW fluctuations.
Since the $d_{z^2}$-orbital component in the CDW fluctuations is
dominant and $\q$ dependence is similar to the SDW fluctuations \cite{bilayer-Inoue}, the contribution
from the CDW fluctuations to $d_{x^2-y^2}$-orbital cold spots is
expected to be small.
Thus, the CDW fluctuations are
expected to make no qualitative difference for the transport phenomena.
In the present study, we ignore the CDW fluctuations for simplicity.

Finally, we comment on the effect of current vertex corrections (CVC) on the transport phenomena. In single-band cuprates, $R_H$ is strongly enhanced by the CVC at low temperatures, since the total current is strongly affected by the CVC \cite{Non-Fermi,Non-Fermi2,Non-Fermi-Cu-RH,Non-Fermi-Cu-RH2,Non-Fermi-FeSeS,Non-Fermi-Fe}. However, the enhancement of $R_H$ due to the CVC is qualitatively different from that arising from qQM.
In multiband systems such as La$_3$Ni$_2$O$_7$, the role of CVC remains unclear and is an important issue for future study.

%%%%%%%%%%%%%%%%%%%%%%%%%%%%%%%%%%%%%%%
%{\it Summary:} \ 
%%%%%%%%%%%%%%%%%%%%%%%%%%%%%%%%%%%%%%%
\section{Summary}

%\subsection{Origin of CDW formation, CDW+SDW fluctuation-mediated SC}
In this paper, we studied the origin of non-Fermi-liquid transport phenomena in
 thin-film bilayer nickelate La$_3$Ni$_2$O$_7$. 
We derived a rigorous formula for the $R_H$ that incorporates the
  important many-body effects within the qQM term. We found that the $T$ dependence of
 the quasiparticle damping $\gamma_b\gg T$ introduced into the qQM term plays an important role in the $T$ dependence of the transport phenomena.  We also found that the $T$ dependence of $R_H$ in La$_3$Ni$_2$O$_7$ is significant due to the competition between the positive contribution from the hole band and the negative one from the electron band.
  The obtained $\rho\propto T$ and positive $R_H$ increasing at low $T$ are consistent with experiments.
  The qQM term is also essential for the Nernst coefficient and other transport phenomena involving the second derivative velocity $v^{\mu\nu}$.

%\subsection{Acknowledgments}
\acknowledgements
We are grateful to Y. Nomura and M. Osada for fruitful discussions.
This work was supported by JSPS KAKENHI Grant Numbers JP25H01246, JP25H01248, JP24K00568, JP24K06938, JP23K03299, JP22K14003.

{\bf Data availability}

The data that support the findings of this article are not publicly available. The data are available from the authors upon reasonable request.

\subsection{Appendix A: Details of the tight-binding model for thin-film La$_3$Ni$_2$O$_7$}
Here, we describe the details of the tight-binding model used for thin-film La$_3$Ni$_2$O$_7$. 
We start with the bilayer 2-orbital bulk tight-binding model based on first-principles calculations under high pressure, as presented in previous work \cite{bilayer-Inoue}. This model includes the $d_{z^2}$ and $d_{x^2-y^2}$ orbitals in each layer.
To reproduce the band
structure near the Fermi energy ($|E|\lesssim 0.5$eV) observed in ARPES measurements of thin-film La$_3$Ni$_2$O$_7$ \cite{Ni-thin-ARPES, Ni-thin-ARPES2}, we modify the bulk model.
Specifically, to reduce the separation between the two $\beta$ FSs facing each other across the X point, while keeping them parallel, we decrease the inter-layer hopping of the $d_{z^2}$ orbital by $28$\%, shift the $d_{z^2}$-orbital level by $-0.45$eV,
and introduce $20$\% hole doping to each Ni atom. 
This reduction in inter-layer hopping is attributed to the increased inter-layer distance in thin-film La$_3$Ni$_2$O$_7$ compared to bulk at high pressure. DFT$+U$ calculation \cite{DFT+U} has reported a reduction of the $d_{z^2}$ interlayer hopping by approximately 30\%, a shift of the $d_{z^2}$-orbital level by $-0.15$ eV, and the presence of 20\% hole doping.
The present modified model is consistent with the DFT$+U$ results except for the magnitude of the $d_{z^2}$-orbital level shift.
In the DFT$+U$ model, both the separation between the two $\beta$ FSs and the size of $\gamma$ FS are larger than those observed in ARPES measurements. Thus, a larger magnitude of the $d_{z^2}$-orbital level shift, as adopted in the present model, is required to reproduce the ARPES measurements.
\color{black}

\subsection{Appendix B: Formulation of FLEX and transport phenomena}
The self-energy in the FLEX approximation $\hat{\Sigma}$, the effective interaction
for the self-energy in the FLEX approximation $\hat{V}$, the spin
(charge) susceptibility $\hat{\chi}^{s(c)}$, the irreducible susceptibility
$\hat{\chi}^0$, and Green's function $\hat{G}$ are given as
\begin{eqnarray}
{\Sigma}_{lm}(k)&=&\frac{T}{N}\sum_{q}V_{ll',mm'}(q)G_{l'm'}(k-q), \label{Eq1}\\
  \hat{V}(q)&=&\frac{3}{2}\hat{\Gamma}^s\hat{\chi}^s(q)\hat{\Gamma}^s+\frac{1}{2}\hat{\Gamma}^c\hat{\chi}^c(q)\hat{\Gamma}^c\label{Sigma}
   \nonumber \\
&& -\frac12 \bigl[{\hat \Gamma}^c{\hat\chi}^0(q){\hat \Gamma}^c
+{\hat \Gamma}^s{\hat\chi}^0(q){\hat \Gamma}^s \nonumber \\
&&-\frac14 ({\hat \Gamma}^s+{\hat \Gamma}^c){\hat\chi}^0(q)
({\hat \Gamma}^s+{\hat \Gamma}^c) \bigr],\\
{\hat{\chi}}^{s(c)}(q)&=&\frac{\hat{\chi}^0(q)}{1-\hat{\Gamma}^{s(c)}{\hat{\chi}}^0(q)}, \\
\chi^0_{ll',mm'}(q)&=&-\frac{T}{N}\sum_kG_{lm}(k+q)G_{m'l'}(k), \\
\hat{G}(k)&=&\left[\frac{1}{i\e_n\hat{1}-\hat{h}_{\k}-\hat{\Sigma}(k)}\right], \label{Eq5}
\end{eqnarray}
where $k=[\k,\e_n=(2n+1)\pi T]$, $q=(\q,\w_m=2m\pi T)$. We put the chemical potential $\mu=0$.
The on-site bare Coulomb interaction ${\hat \Gamma}^s$ for the spin
channel on the upper layer ($l_1,l_2,l_3,l_4=1,2$) or the lower layer ($l_1,l_2,l_3,l_4=3,4$) is given as
\begin{equation}
(\Gamma^{\mathrm{s}})_{l_{1}l_{2},l_{3}l_{4}} = \begin{cases}
U, & l_1=l_2=l_3=l_4 \\
U' , & l_1=l_3 \neq l_2=l_4 \\
J, & l_1=l_2 \neq l_3=l_4 \\
J, & l_1=l_4 \neq l_2=l_3 \\
0 , & \mathrm{otherwise},
\end{cases}
\end{equation}
where the inter-orbital Coulomb interaction $U'=U-2J$.
Also, the on-site bare Coulomb interaction ${\hat \Gamma}^c$ for the charge channel is given as
\begin{equation}
({\hat \Gamma}^{\mathrm{c}})_{l_{1}l_{2},l_{3}l_{4}} = \begin{cases}
-U, & l_1=l_2=l_3=l_4 \\
U'-2J, & l_1=l_3 \neq l_2=l_4 \\
-2U' + J, & l_1=l_2 \neq l_3=l_4 \\
-J, &l_1=l_4 \neq l_2=l_3 \\
0 . & \mathrm{otherwise}.
\end{cases}
\end{equation}

In the present
3D calculation, we employ
$N=128\times128$ $\k(\q)$ meshes and 512 Matsubara
frequencies. ${\hat{h}}_{\k}$ is the matrix expression of the Hamiltonian $H$.

To obtain the real frequency dependence, we use the analytic
continuation by the Pade approximation,
 \begin{eqnarray}
  \hat{\Sigma}(\k,i\e_n)&\rightarrow&  \hat{\Sigma}^{\rm R}_{\k}(\e),\\
%     \hat{V}(\k,i\e_n)&\rightarrow&  \hat{V}^R(\k,\omega),\\
  \hat{G}^{\rm R}_{\bm{k}}(\epsilon)&=&\left[\frac{1}{\epsilon\hat{1}-\hat{h}_{\k}-\hat{\Sigma}^{\rm R}_{\bm{k}}(\epsilon)}\right].
    \end{eqnarray}

 The conductivity $\sigma_{xx}$ is given as,
 \begin{eqnarray}
\sigma_{xx} =  e^2 &&\sum_{\bm{k},a}\int\frac{d\epsilon}{\pi}\ 
 \left(-\frac{\partial f}{\partial \epsilon}\right)\\ \nonumber
&&  \times v^x_a
  G^{\rm R}_av^x_aG^{\rm A}_a,
 \end{eqnarray}
 where $f(\epsilon)=[\exp(\epsilon/T)+1]^{-1}$, and $-e$ $(e>0)$ is the electron charge. The velocity is given as $v^x_{a,\k}=\frac{\partial
\epsilon_{a,\k}}{\partial k_x}$, and the resistivity along $x$-axis is given as,
\begin{equation}
 \rho=\frac{1}{\sigma_{xx}}.
\end{equation}

The Seebeck coefficient $S$ is given by
\begin{eqnarray}
 S&=&-e\frac{\alpha_{xx}}{T\sigma_{xx}},\\
\alpha_{xx} &=& \sum_{\bm{k},a}\int\frac{d\epsilon}{\pi}\ 
  \left(-\frac{\partial f}{\partial \epsilon}\right)\epsilon\\ \nonumber 
&&  \times v^x_a
  G^{\rm R}_av^x_aG^{\rm A}_{a}.
  \end{eqnarray}

The Nernst coefficient $\nu$ is given by
\begin{equation}
\nu=\frac{\alpha_{xy}}{B\sigma_{xx}}-\frac{S\sigma_{xy}}{B\sigma_{xx}},
\end{equation}

\begin{eqnarray}
\alpha_{xy}&=&B\frac{e^2}{T}\sum_{\k,a}\int\frac{d\epsilon}{\pi}f'(\e)\epsilon|{\rm Im}G_a^{\rm R}||G_a^{\rm R}|^2\nonumber\\
 & &\times v^x_{a}[v^y_a v_a^{xy}(\e) - v^x_a v_a^{yy}(\e)]\nonumber\\
% &=& B\frac{e^2}{T} \sum_{\k,a} f'(\e_a) \e_a v^x_{a}[v^y_a
 % v_a^{xy}(\e_a) - v^x_a v_a^{yy}(\e_a)]\frac1{2\gamma_a^2}. \label{eqn:Ner}
\end{eqnarray}

In order to convert the unit of transport
coefficients to the experimental unit, we use lattice
constant $a=b=3.715$\r{A}, $c=19.7$\r{A}.

\subsection{Appendix C: Additional results}

\begin{figure}[!htb]
	\includegraphics[width=.99\linewidth]{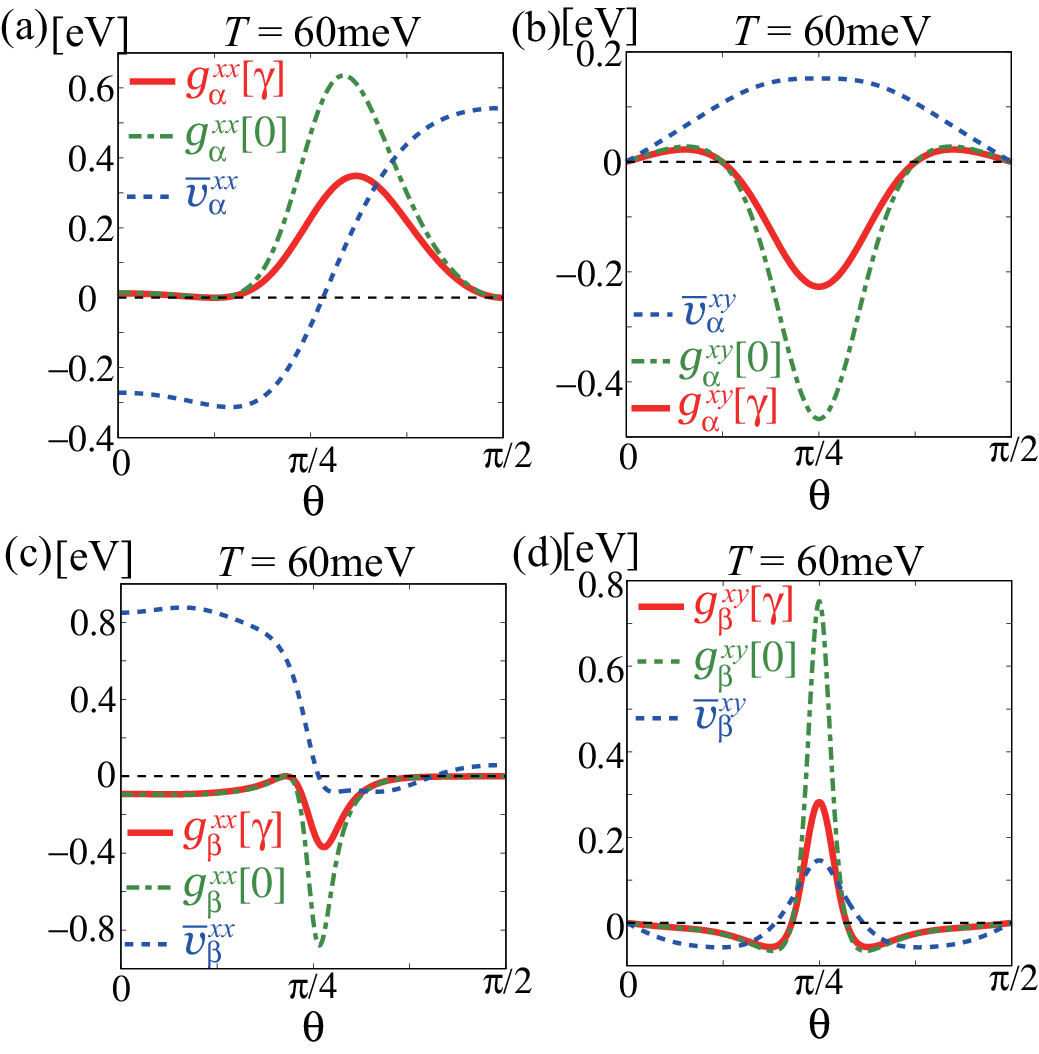}
	\caption{
 $\theta$ dependences of (a) $g_\a^{xx}[\gamma]$, $g_\a^{xx}[0]$, and
 $\bar{v}_\a^{xx}$, (b) $g_\a^{xy}[\gamma]$, $g_\a^{xy}[0]$, and
 $\bar{v}_\a^{xy}$, (c) $g_\b^{xx}[\gamma]$, $g_\b^{xx}[0]$, and
 $\bar{v}_\b^{xx}$, and (d) $g_\b^{xy}[\gamma]$, $g_\b^{xy}[0]$, and
 $\bar{v}_\b^{xy}$ on FSs $(\e=\mu)$ for $U=2.74$eV, $J/U=0.2$ at $T=60$meV. $\theta=\pi/4$ corresponds
 to the cold spot.
}
\label{fig:fig13}
\end{figure}

Figure \ref{fig:fig13} shows $\theta$ dependences of
$g^{\mu\nu}_a[\gamma]$, $g^{\mu\nu}_a[0]$, and $\bar{v}^{\mu\nu}_a$ on
FSs $(\e=\mu)$ for
$U=2.74$eV, $J/U=0.2$
 at $T=60$meV. The value of $g^{\mu\nu}_a[\gamma]$ at cold spot
 $(\theta=\pi/4)$ is smaller than that of
 $g^{\mu\nu}_a[0]$, since the quasiparticle damping is large at high temperatures. 
Due to the suppression of $g^{\mu\nu}_a[\gamma]$, the value of $R_H$ is close to that of $R_H^{\rm no \, qQM}$ at high temperatures, as shown in Fig. \ref{fig:fig3}. In contrast,  $g^{\mu\nu}_a[\gamma]\sim g^{\mu\nu}_a[0]$ is satisfied at low temperatures. Thus, the value of $R_H$ is close to that of $R_H^{\rm RTA}$ at low temperatures.

%%%%%%%%%%%%%%%%%%%%%%%%%%%%%%%%%%%%%%%
%%%%%%%%%%%%%%%%%%%%%%%%%%%%%%%%%%%%%%

\begin{figure}[!htb]
	\includegraphics[width=.99\linewidth]{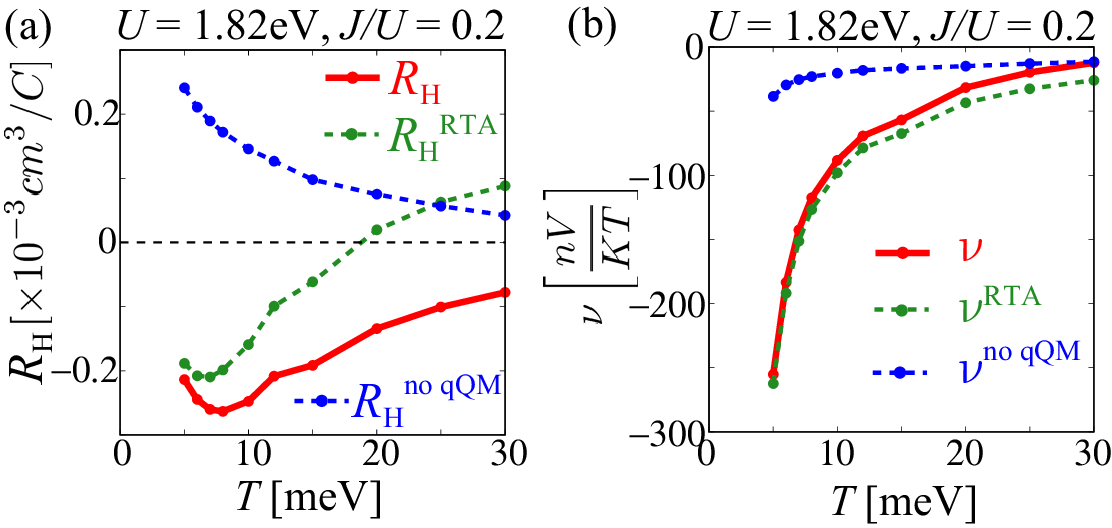}
	\caption{
(a) $T$ dependences of $R_H$, $R_H^{\rm RTA}$, and $R_H^{\rm no \, qQM}$ for $U=1.82$eV, $J/U=0.2$
 ($\alpha_s=0.93$ at $T=10$meV), and (b) Those of $\nu$, $\nu^{\rm RTA}$, and $\nu^{\rm no \, qQM}$
}
\label{fig:fig4}
\end{figure}

Figures \ref{fig:fig4} (a) and (b) exhibit $T$ dependences of $R_H$ and
$\nu$ for $U=1.82$eV, $J/U=0.2$. Here, $\alpha_s=0.93$ is satisfied at $T=10$meV. In this case, $R_H$ and $\nu$ are negative at low $T$. The $T$ dependence of $R_H$ is
sensitive to the competition between the contributions from $\alpha$ and
$\beta$ pockets, whereas the negative $\nu$ decreasing at low $T$ is
robust in the thin-film La$_3$Ni$_2$O$_7$ model.
We find that $R_H$ becomes negative in the regime of weak
electronic correlations, i.e., when the values of $U$ and $J$ are small.
$R_H$ and $\nu$ are closer to $R_H^{\rm RTA}$ and
$\nu^{\rm RTA}$ than Figs. \ref{fig:fig3} (a) and (d) for
$U=2.74$eV, since $\gamma$ for $U=1.82$eV is smaller than that for $U= 2.74$eV.

%%%%%%%%%%%%%%%%%%%%%%%%%%%%%%%%%%%%%%%
%%%%%%%%%%%%%%%%%%%%%%%%%%%%%%%%%%%%%%

\begin{figure}[!htb]
	\includegraphics[width=.99\linewidth]{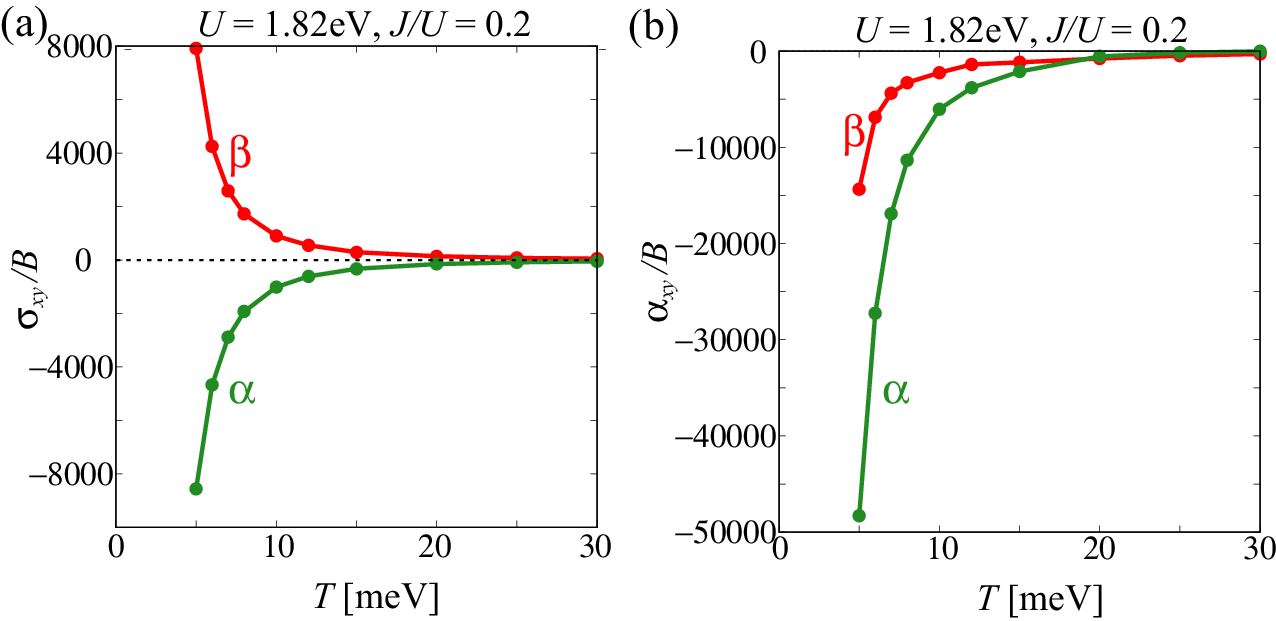}
	\caption{
 $T$ dependences of (a) $\sigma_{xy}/B$ and (b) $\alpha_{xy}/B$ for
 $\alpha$ and $\beta$ bands given by using
 $g^{\mu\nu}_a[\gamma]$ for $U=1.82$eV with $J/U=0.2$.
}
\label{fig:fig5-2}
\end{figure}

Figures \ref{fig:fig5-2} (a) and (b) show the contributions of $\sigma_{xy}/B$ and
$\alpha_{xy}/B$ from $\alpha$ and $\beta$ bands given by using
 $g^{\mu\nu}_a[\gamma]$ for $U=1.82$eV, $J/U=0.2$.
 These behaviors are similar to those for $U=1.82$eV, $J/U=0.2$ in
 Figs. \ref{fig:fig5}(a) and (b).
 
%%%%%%%%%%%%%%%%%%%%%%%%%%%%%%%%%%%%%%%
%%%%%%%%%%%%%%%%%%%%%%%%%%%%%%%%%%%%%%

\color{black}

\begin{figure}[!htb]
	\includegraphics[width=.60\linewidth]{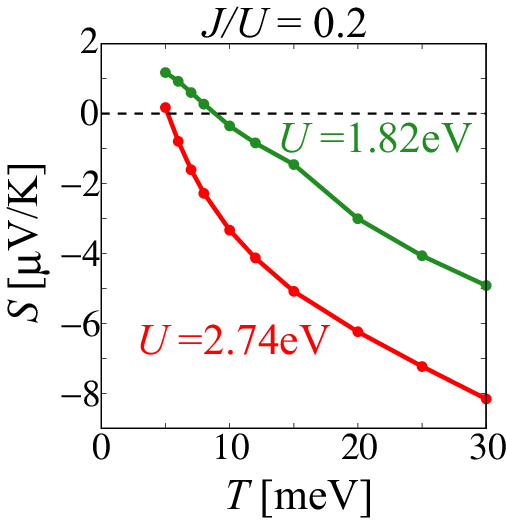}
	\caption{
$T$ dependences of Seebeck coefficient $S$ for $U=2.74$eV and
 $1.82$eV at $J/U=0.2$.
}
\label{fig:figS6}
\end{figure}

Figure \ref{fig:figS6} shows $T$ dependences of $S$ for $U=2.74$eV and
$U=1.82$eV at $J/U=0.2$. The obtained $S$ is negative and increases with decreasing $T$, which is
similar to the behavior observed in triple-layer and quintuple-layer
nickelates \cite{Ni-S}.

%%%%%%%%%%%%%%%%%%%%%%%%%%%%%%%%%%%%%%%

%%%%%%%%%%%%%%%%%%%%%%%%
%references
%%%%%%%%%%%%%%%%%%%%%%%%

\end{document}